\documentclass{aastex631}

\pdfoutput=1

\usepackage{amsmath}
\usepackage{threeparttablex}
\usepackage{multirow}
\usepackage{xspace}
\usepackage{placeins}

\newcommand{\nclusters}{65}
\newcommand{\nsources}{25}
\newcommand{\nappsources}{14}
\newcommand{\fitburst}{\texttt{fitburst}}

\begin{document}
\title{CHIME/FRB Discovery of \nsources\ Repeating Fast Radio Burst Sources}
\shorttitle{CHIME/FRB Discovery of \nsources\ Repeaters}
\shortauthors{The CHIME/FRB Collaboration}

\author[0000-0001-5908-3152]{Bridget C.~Andersen}
  \affiliation{Department of Physics, McGill University, 3600 rue University, Montr\'eal, QC H3A 2T8, Canada}
  \affiliation{Trottier Space Institute, McGill University, 3550 rue University, Montr\'eal, QC H3A 2A7, Canada}
\author[0000-0003-3772-2798]{Kevin Bandura}
  \affiliation{Lane Department of Computer Science and Electrical Engineering, 1220 Evansdale Drive, PO Box 6109 Morgantown, WV 26506, USA}
  \affiliation{Center for Gravitational Waves and Cosmology, West Virginia University, Chestnut Ridge Research Building, Morgantown, WV 26505, USA}
\author[0000-0002-3615-3514]{Mohit Bhardwaj}
  \affiliation{Department of Physics, McGill University, 3600 rue University, Montr\'eal, QC H3A 2T8, Canada}
  \affiliation{Trottier Space Institute, McGill University, 3550 rue University, Montr\'eal, QC H3A 2A7, Canada}
  \affiliation{Department of Physics, Carnegie Mellon University, 5000 Forbes Avenue, Pittsburgh, 15213, PA, USA}
\author[0000-0001-8537-9299]{P.~J.~Boyle}
  \affiliation{Department of Physics, McGill University, 3600 rue University, Montr\'eal, QC H3A 2T8, Canada}
  \affiliation{Trottier Space Institute, McGill University, 3550 rue University, Montr\'eal, QC H3A 2A7, Canada}
\author[0000-0002-1800-8233]{Charanjot Brar}
  \affiliation{Department of Physics, McGill University, 3600 rue University, Montr\'eal, QC H3A 2T8, Canada}
  \affiliation{Trottier Space Institute, McGill University, 3550 rue University, Montr\'eal, QC H3A 2A7, Canada}
\author[0000-0003-2047-5276]{Tomas Cassanelli}
  \affiliation{Department of Electrical Engineering, Universidad de Chile, Av. Tupper 2007, Santiago 8370451, Chile}
\author[0000-0002-2878-1502]{S.~Chatterjee}
  \affiliation{Department of Astronomy and Cornell Center for Astrophysics and Planetary Science, Cornell University, Ithaca NY 14853, USA}
\author[0000-0002-3426-7606]{Pragya Chawla}
  \affiliation{Anton Pannekoek Institute for Astronomy, University of Amsterdam, Science Park 904, 1098 XH Amsterdam, The Netherlands}
\author[0000-0001-6422-8125]{Amanda M.~Cook}
  \affiliation{Dunlap Institute for Astronomy \& Astrophysics, University of Toronto, 50 St.~George Street, Toronto, ON M5S 3H4, Canada}
  \affiliation{David A.~Dunlap Department of Astronomy \& Astrophysics, University of Toronto, 50 St.~George Street, Toronto, ON M5S 3H4, Canada}
\author[0000-0002-8376-1563]{Alice P.~Curtin}
  \affiliation{Department of Physics, McGill University, 3600 rue University, Montr\'eal, QC H3A 2T8, Canada}
  \affiliation{Trottier Space Institute, McGill University, 3550 rue University, Montr\'eal, QC H3A 2A7, Canada}
\author[0000-0001-7166-6422]{Matt Dobbs}
  \affiliation{Department of Physics, McGill University, 3600 rue University, Montr\'eal, QC H3A 2T8, Canada}
  \affiliation{Trottier Space Institute, McGill University, 3550 rue University, Montr\'eal, QC H3A 2A7, Canada}
\author[0000-0003-4098-5222]{Fengqiu Adam Dong}
  \affiliation{Department of Physics and Astronomy, University of British Columbia, 6224 Agricultural Road, Vancouver, BC V6T 1Z1 Canada}
\author[0000-0001-9855-5781]{Jakob T. Faber}
  \affiliation{Department of Physics, McGill University, 3600 rue University, Montr\'eal, QC H3A 2T8, Canada}
  \affiliation{Cahill Center for Astronomy and Astrophysics, MC 249-17 California Institute of Technology, Pasadena CA 91125, USA}
\author[0000-0002-6899-1176]{Mateus Fandino}
  \affiliation{Department of Physics and Astronomy, University of British Columbia, 6224 Agricultural Road, Vancouver, BC V6T 1Z1 Canada}
  \affiliation{Department of Physical Sciences, Thompson Rivers University, Kamloops, BC, Canada}
\author[0000-0001-8384-5049]{Emmanuel Fonseca}
  \affiliation{Department of Physics and Astronomy, West Virginia University, P.O. Box 6315, Morgantown, WV 26506, USA }
  \affiliation{Center for Gravitational Waves and Cosmology, West Virginia University, Chestnut Ridge Research Building, Morgantown, WV 26505, USA}
\author[0000-0002-3382-9558]{B.~M.~Gaensler}
  \affiliation{Dunlap Institute for Astronomy \& Astrophysics, University of Toronto, 50 St.~George Street, Toronto, ON M5S 3H4, Canada}
  \affiliation{David A.~Dunlap Department of Astronomy \& Astrophysics, University of Toronto, 50 St.~George Street, Toronto, ON M5S 3H4, Canada}
\author[0000-0001-5553-9167]{Utkarsh Giri}
  \affiliation{Department of Physics, University of Wisconsin-Madison, Madison, WI 53706, USA}
\author[0000-0002-3654-4662]{Antonio Herrera-Martin}
  \affiliation{David A.~Dunlap Department of Astronomy \& Astrophysics, University of Toronto, 50 St.~George Street, Toronto, ON M5S 3H4, Canada}
\author[0000-0001-7301-5666]{Alex S.~Hill}
  \affiliation{Department of Computer Science, Math, Physics, \& Statistics, University of British Columbia, Kelowna, BC V1V 1V7, Canada}
  \affiliation{Dominion Radio Astrophysical Observatory, Herzberg Research Centre for Astronomy and Astrophysics, National Research Council Canada, PO Box 248, Penticton, BC V2A 6J9, Canada}
\author[0000-0003-2405-2967]{Adaeze Ibik}
  \affiliation{Dunlap Institute for Astronomy \& Astrophysics, University of Toronto, 50 St.~George Street, Toronto, ON M5S 3H4, Canada}
  \affiliation{David A.~Dunlap Department of Astronomy \& Astrophysics, University of Toronto, 50 St.~George Street, Toronto, ON M5S 3H4, Canada}
\author[0000-0003-3059-6223]{Alexander Josephy}
  \affiliation{Department of Physics, McGill University, 3600 rue University, Montr\'eal, QC H3A 2T8, Canada}
  \affiliation{Trottier Space Institute, McGill University, 3550 rue University, Montr\'eal, QC H3A 2A7, Canada}
\author[0000-0003-4810-7803]{Jane F.~Kaczmarek}
  \affiliation{Dominion Radio Astrophysical Observatory, Herzberg Research Centre for Astronomy and Astrophysics, National Research Council Canada, PO Box 248, Penticton, BC V2A 6J9, Canada}
\author[0000-0003-2739-5869]{Zarif Kader}
  \affiliation{Department of Physics, McGill University, 3600 rue University, Montr\'eal, QC H3A 2T8, Canada}
  \affiliation{Trottier Space Institute, McGill University, 3550 rue University, Montr\'eal, QC H3A 2A7, Canada}
\author[0000-0001-9345-0307]{Victoria Kaspi}
  \affiliation{Department of Physics, McGill University, 3600 rue University, Montr\'eal, QC H3A 2T8, Canada}
  \affiliation{Trottier Space Institute, McGill University, 3550 rue University, Montr\'eal, QC H3A 2A7, Canada}
\author[0000-0003-1455-2546]{T.~L.~Landecker}
  \affiliation{Dominion Radio Astrophysical Observatory, Herzberg Research Centre for Astronomy and Astrophysics, National Research Council Canada, PO Box 248, Penticton, BC V2A 6J9, Canada}
\author[0000-0003-2116-3573]{Adam E.~Lanman}
  \affiliation{Department of Physics, McGill University, 3600 rue University, Montr\'eal, QC H3A 2T8, Canada}
  \affiliation{Trottier Space Institute, McGill University, 3550 rue University, Montr\'eal, QC H3A 2A7, Canada}
\author[0000-0002-5857-4264]{Mattias Lazda}
  \affiliation{Department of Physics, McGill University, 3600 rue University, Montr\'eal, QC H3A 2T8, Canada}
\author[0000-0002-4209-7408]{Calvin Leung}
  \affiliation{MIT Kavli Institute for Astrophysics and Space Research, Massachusetts Institute of Technology, 77 Massachusetts Ave, Cambridge, MA 02139, USA}
  \affiliation{Department of Physics, Massachusetts Institute of Technology, 77 Massachusetts Ave, Cambridge, MA 02139, USA}
\author[0000-0001-7453-4273]{Hsiu-Hsien Lin}
  \affiliation{Institute of Astronomy and Astrophysics, Academia Sinica, Astronomy-Mathematics Building, No. 1, Sec. 4, Roosevelt Road, Taipei 10617, Taiwan}
  \affiliation{Canadian Institute for Theoretical Astrophysics, 60 St.~George Street, Toronto, ON M5S 3H8, Canada}
\author[0000-0002-4279-6946]{Kiyoshi W.~Masui}
  \affiliation{MIT Kavli Institute for Astrophysics and Space Research, Massachusetts Institute of Technology, 77 Massachusetts Ave, Cambridge, MA 02139, USA}
  \affiliation{Department of Physics, Massachusetts Institute of Technology, 77 Massachusetts Ave, Cambridge, MA 02139, USA}
\author[0000-0001-7348-6900]{Ryan Mckinven}
  \affiliation{Department of Physics, McGill University, 3600 rue University, Montr\'eal, QC H3A 2T8, Canada}
  \affiliation{Trottier Space Institute, McGill University, 3550 rue University, Montr\'eal, QC H3A 2A7, Canada}
\author[0000-0002-0772-9326]{Juan Mena-Parra}
  \affiliation{Dunlap Institute for Astronomy \& Astrophysics, University of Toronto, 50 St.~George Street, Toronto, ON M5S 3H4, Canada}
  \affiliation{David A.~Dunlap Department of Astronomy \& Astrophysics, University of Toronto, 50 St.~George Street, Toronto, ON M5S 3H4, Canada}
\author[0000-0001-8845-1225]{Bradley W.~Meyers}
  \affiliation{International Centre for Radio Astronomy Research (ICRAR), Curtin University, Bentley WA 6102 Australia}
\author[0000-0002-2551-7554]{D.~Michilli}
  \affiliation{MIT Kavli Institute for Astrophysics and Space Research, Massachusetts Institute of Technology, 77 Massachusetts Ave, Cambridge, MA 02139, USA}
  \affiliation{Department of Physics, Massachusetts Institute of Technology, 77 Massachusetts Ave, Cambridge, MA 02139, USA}
\author[0000-0002-3616-5160]{Cherry Ng}
  \affiliation{Dunlap Institute for Astronomy \& Astrophysics, University of Toronto, 50 St.~George Street, Toronto, ON M5S 3H4, Canada}
\author[0000-0002-8897-1973]{Ayush Pandhi}
  \affiliation{Dunlap Institute for Astronomy \& Astrophysics, University of Toronto, 50 St.~George Street, Toronto, ON M5S 3H4, Canada}
  \affiliation{David A.~Dunlap Department of Astronomy \& Astrophysics, University of Toronto, 50 St.~George Street, Toronto, ON M5S 3H4, Canada}
\author[0000-0002-8912-0732]{Aaron B.~Pearlman}
  \affiliation{Department of Physics, McGill University, 3600 rue University, Montr\'eal, QC H3A 2T8, Canada}
  \affiliation{Trottier Space Institute, McGill University, 3550 rue University, Montr\'eal, QC H3A 2A7, Canada}
\author[0000-0003-2155-9578]{Ue-Li Pen}
  \affiliation{Institute of Astronomy and Astrophysics, Academia Sinica, Astronomy-Mathematics Building, No. 1, Sec. 4, Roosevelt Road, Taipei 10617, Taiwan}
  \affiliation{Canadian Institute for Theoretical Astrophysics, 60 St.~George Street, Toronto, ON M5S 3H8, Canada}
  \affiliation{Canadian Institute for Advanced Research, MaRS Centre, West Tower, 661 University Avenue, Suite 505 }
  \affiliation{Dunlap Institute for Astronomy \& Astrophysics, University of Toronto, 50 St.~George Street, Toronto, ON M5S 3H4, Canada}
  \affiliation{Perimeter Institute for Theoretical Physics, 31 Caroline Street N, Waterloo, ON N25 2YL, Canada}
\author[0000-0002-9822-8008]{Emily Petroff}
  \affiliation{Department of Physics, McGill University, 3600 rue University, Montr\'eal, QC H3A 2T8, Canada}
  \affiliation{Trottier Space Institute, McGill University, 3550 rue University, Montr\'eal, QC H3A 2A7, Canada}
\author[0000-0002-4795-697X]{Ziggy Pleunis}
  \affiliation{Dunlap Institute for Astronomy \& Astrophysics, University of Toronto, 50 St.~George Street, Toronto, ON M5S 3H4, Canada}
\author[0000-0001-7694-6650]{Masoud Rafiei-Ravandi}
  \affiliation{Department of Physics, McGill University, 3600 rue University, Montr\'eal, QC H3A 2T8, Canada}
  \affiliation{Trottier Space Institute, McGill University, 3550 rue University, Montr\'eal, QC H3A 2A7, Canada}
\author[0000-0003-1842-6096]{Mubdi Rahman}
  \affiliation{Sidrat Research, 124 Merton Street, Toronto, ON M4S 2Z2, Canada}
\author[0000-0001-5799-9714]{Scott M.~Ransom}
  \affiliation{National Radio Astronomy Observatory, 520 Edgemont Rd, Charlottesville, VA 22903, USA}
\author[0000-0003-3463-7918]{Andre Renard}
  \affiliation{Dunlap Institute for Astronomy \& Astrophysics, University of Toronto, 50 St.~George Street, Toronto, ON M5S 3H4, Canada}
\author[0000-0003-3154-3676]{Ketan R.~Sand}
  \affiliation{Department of Physics, McGill University, 3600 rue University, Montr\'eal, QC H3A 2T8, Canada}
  \affiliation{Trottier Space Institute, McGill University, 3550 rue University, Montr\'eal, QC H3A 2A7, Canada}
\author[0000-0001-5504-229X]{Pranav Sanghavi}
  \affiliation{Department of Physics, Yale University, New Haven, CT 06520, USA}
\author[0000-0002-7374-7119]{Paul Scholz}
  \affiliation{Dunlap Institute for Astronomy \& Astrophysics, University of Toronto, 50 St.~George Street, Toronto, ON M5S 3H4, Canada}
\author[0000-0002-4823-1946]{Vishwangi Shah}
  \affiliation{Department of Physics, McGill University, 3600 rue University, Montr\'eal, QC H3A 2T8, Canada}
  \affiliation{Trottier Space Institute, McGill University, 3550 rue University, Montr\'eal, QC H3A 2A7, Canada}
\author[0000-0002-6823-2073]{Kaitlyn Shin}
  \affiliation{MIT Kavli Institute for Astrophysics and Space Research, Massachusetts Institute of Technology, 77 Massachusetts Ave, Cambridge, MA 02139, USA}
  \affiliation{Department of Physics, Massachusetts Institute of Technology, 77 Massachusetts Ave, Cambridge, MA 02139, USA}
\author[0000-0003-2631-6217]{Seth Siegel}
  \affiliation{Department of Physics, McGill University, 3600 rue University, Montr\'eal, QC H3A 2T8, Canada}
  \affiliation{Perimeter Institute for Theoretical Physics, 31 Caroline Street N, Waterloo, ON N25 2YL, Canada}
\author[0000-0002-2088-3125]{Kendrick Smith}
  \affiliation{Perimeter Institute for Theoretical Physics, 31 Caroline Street N, Waterloo, ON N25 2YL, Canada}
\author[0000-0001-9784-8670]{Ingrid Stairs}
  \affiliation{Department of Physics and Astronomy, University of British Columbia, 6224 Agricultural Road, Vancouver, BC V6T 1Z1 Canada}
\author[0000-0003-0607-8194]{Jianing Su}
  \affiliation{Department of Physics, McGill University, 3600 rue University, Montr\'eal, QC H3A 2T8, Canada}
\author[0000-0003-2548-2926]{Shriharsh P.~Tendulkar}
  \affiliation{Department of Astronomy and Astrophysics, Tata Institute of Fundamental Research, Mumbai, 400005, India}
  \affiliation{National Centre for Radio Astrophysics, Post Bag 3, Ganeshkhind, Pune, 411007, India}
\author[0000-0003-4535-9378]{Keith Vanderlinde}
  \affiliation{Dunlap Institute for Astronomy \& Astrophysics, University of Toronto, 50 St.~George Street, Toronto, ON M5S 3H4, Canada}
  \affiliation{David A.~Dunlap Department of Astronomy \& Astrophysics, University of Toronto, 50 St.~George Street, Toronto, ON M5S 3H4, Canada}
\author[0000-0002-1491-3738]{Haochen Wang}
  \affiliation{MIT Kavli Institute for Astrophysics and Space Research, Massachusetts Institute of Technology, 77 Massachusetts Ave, Cambridge, MA 02139, USA}
  \affiliation{Department of Physics, Massachusetts Institute of Technology, 77 Massachusetts Ave, Cambridge, MA 02139, USA}
\author[0000-0001-7314-9496]{Dallas Wulf}
  \affiliation{Department of Physics, McGill University, 3600 rue University, Montr\'eal, QC H3A 2T8, Canada}
  \affiliation{Trottier Space Institute, McGill University, 3550 rue University, Montr\'eal, QC H3A 2A7, Canada}
\author[0000-0001-8278-1936]{Andrew Zwaniga}
  \affiliation{Department of Physics, Toronto Metropolitan University, 350 Victoria St, Toronto, ON M5B 2K3, Canada}
\newcommand{\allacks}{
FRB research at WVU is supported by an NSF grant (2006548, 2018490).
%
%
%
FRB research at UBC is supported by an NSERC Discovery Grant and by the Canadian Institute for Advanced Research. The CHIME/FRB baseband system is funded in part by a Canada Foundation for Innovation John R. Evans Leaders Fund award to IHS.

B.C.A. is supported by an FRQNT Doctoral Research Award.
M.B. is a McWilliams Fellow.
A.M.C. is supported by an NSERC Doctoral Postgraduate Scholarship. 
A.P.C is a Vanier Canada Graduate Scholar.
M.D. is supported by a CRC Chair, NSERC Discovery Grant, CIFAR, and by the FRQNT Centre de Recherche en Astrophysique du Qu\'ebec (CRAQ).
F.A.D. is funded by the UBC Four Year Fellowship.
J.T.F. is a Fulbright Scholar.
B.M.G. acknowledges the support of the Natural Sciences and Engineering Research Council of Canada (NSERC) through grant RGPIN-2022-03163, and of the Canada Research Chairs program.
V.M.K. holds the Lorne Trottier Chair in Astrophysics \& Cosmology, a Distinguished James McGill Professorship, and receives support from an NSERC Discovery grant (RGPIN 228738-13), from an R. Howard Webster Foundation Fellowship from CIFAR, and from the FRQNT CRAQ.
C.L. was supported by the U.S. Department of Defense (DoD) through the National Defense Science \& Engineering Graduate Fellowship (NDSEG) Program.
K.W.M. holds the Adam J. Burgasser Chair in Astrophysics and is supported by an NSF Grant (2008031).
A.P. is supported by the Ontario Graduate Scholarship.
A.B.P. is a Trottier Space Institute (TSI) Fellow and a Fonds de Recherche du Quebec - Nature et Technologies (FRQNT) postdoctoral fellow.
U.L.P. receives the support of the Natural Sciences and Engineering Research Council of Canada (NSERC), [funding reference number RGPIN-2019-067, CRD 523638-18, 555585-20], Ontario Research Fund—research Excellence Program (ORF-RE), Canadian Institute for Advanced Research (CIFAR), Thoth Technology Inc, Alexander von Humboldt Foundation, and the National Science and Technology Council (NSTC) of Taiwan (111-2123-M-001 -008-, and 111-2811-M-001 -040-). 
Z.P. is a Dunlap Fellow.
S.M.R. is a CIFAR Fellow and is supported by the NSF Physics Frontiers Center awards 1430284 and 2020265.
P.S. is a Dunlap Fellow.
K.S. is supported by the NSF Graduate Research Fellowship Program.
S.P.T. is a CIFAR Azrieli Global Scholar in the Gravity and Extreme Universe Program.
A.Z. is supported by the Ontario Graduate Scholarship.
}
\collaboration{1000}{The CHIME/FRB Collaboration}

\correspondingauthor{Ziggy Pleunis}
\email{ziggy.pleunis@dunlap.utoronto.ca}

\begin{abstract}
We present the discovery of \nsources\ new repeating fast radio burst (FRB) sources found among CHIME/FRB events detected between 2019 September 30 and 2021 May 1.  The sources were found using a new clustering algorithm that looks for multiple events co-located on the sky having similar dispersion measures (DMs). The new repeaters have DMs ranging from $\sim$220~pc~cm$^{-3}$ to $\sim$1700~pc~cm$^{-3}$, and include sources having exhibited as few as two bursts to as many as twelve. We report a statistically significant difference in both the DM and extragalactic DM (eDM) distributions between repeating and apparently nonrepeating sources, with repeaters having lower mean DM and eDM, and we discuss the implications. We find no clear bimodality between the repetition rates of repeaters and upper limits on repetition from apparently nonrepeating sources after correcting for sensitivity and exposure effects, although some active repeating sources stand out as anomalous. We measure the repeater fraction over time and find that it tends to an equilibrium of $2.6_{-2.6}^{+2.9}$\% over our total time-on-sky thus far. We also report on \nappsources\ more sources which are promising repeating FRB candidates and which merit follow-up observations for confirmation.
\end{abstract}


\section{Introduction}
\label{sec:intro}

Fast radio bursts (FRBs) are microsecond--millisecond flashes of radio waves that are detectable over extragalactic distances \citep[see][for reviews of the phenomenon]{cc19,phl19}. Some FRB sources repeat, which rules out cataclysmic models for burst production \citep{ssh+16a}. Magnetars are likely to be the origin of at least a fraction of FRBs as the two source classes have been directly linked \citep{brb+20,abb+20}. The sources that repeat show clustered activity \citep{oyp18} that in some cases is periodic, with bursts being detectable only during a window of time in each cycle \citep{aab+20,rms+20,css+21}. For one source, this periodic activity has been shown to be chromatic, with activity at lower frequencies trailing the activity at higher frequencies \citep{pcl+21,pmb+21,bsm+22}. Importantly, it is as-yet unclear whether all FRBs repeat \citep{csr+19,jof+20a,jof+20b}, what fraction of repeating sources show periodic activity, and on what timescales repeaters are active.

Complex time-frequency structure, with subbursts that drift downward in frequency as time progresses, seems ubiquitous among bursts from repeating sources \citep{hss+19,abb+19b}. On average, repeater bursts are observed to be longer in duration and narrower in bandwidth than for bursts seen from nonrepeating sources \citep{pgk+21}. It is unclear if this observed difference represents an intrinsic difference in the two populations, if it is caused by a propagation effect related to distinct source environments, or if it is due to a selection effect that leads to the observation of two extremes out of a continuum of source properties \citep[e.g.,][]{cmg20}. 

Repeating sources of FRBs provide the best opportunity for detailed follow-up observations, and the most prolific of repeaters -- FRB~20121102A \citep{ssh+16a}, FRB~20180916B \citep{abb+19c}, FRB~20190520B \citep{nal+22}, FRB~20200120E \citep{bgk+21}, and FRB~20201124A \citep{lac+22} -- have provided many insights. All five have been localized to sub-arcsecond precision. They are associated with a star-forming dwarf galaxy, a disk galaxy, another star-forming dwarf galaxy, a globular cluster, and a barred galaxy, respectively \citep{tbc+17,mnh+20,nal+22,kmn+22,xnc+22}, demonstrating that repeaters inhabit a variety of host types. Two sources of repeating FRBs that live in extreme magneto-ionic environments are coincident with persistent radio sources \citep[PRSs;][]{msh+18,nal+22} but for other sources, PRSs of luminosity similar to these two have been ruled out \citep{lca22}. The complex magnetized environments of repeaters can result in depolarization of bursts towards lower frequencies \citep{fly+22}, depolarization due to Faraday conversion \citep{ksl+22} and temporal evolution of the Faraday rotation that can be secular \citep{hms+21}, a combination of stochastic and secular \citep{mgm+22a} or exhibit a sign flip pointing to a magnetic field reversal \citep{acb+22}.

Repeater bursts have been resolved to timescales from a few $\mu$s to a few ms \citep[FRB~20180916B and FRB~20200120E;][]{nhk+21,nhk+21b} and they have been detected up to 8\,GHz \citep[FRB~20121102A;][]{gsp+18} and down to 110\,MHz \citep[FRB~20180916B;][]{pcl+21,pmb+21}. Repeater bursts can be extremely narrowband \citep[65\,MHz;][]{ksf+21} and show order-of-magnitude differences in brightness from subburst to subburst \citep{cab+20,csa+20} or from burst envelope to burst envelope \citep{kso+19}. The unprecedented sensitivity of the Five-hundred-meter Aperture Spherical Telescope enables detailed measurements of burst energy distributions \citep{lwz+21}, which informs possible progenitor models and emission mechanisms through the determination of minimum and maximum characteristic energies.\footnote{\citet{lwz+21} claim a bimodal energy distribution for FRB~20121102A, but \citet{agg21} argues that the bimodality disappears when energies are corrected for bandwidth.} Despite coordinated efforts, a prompt multi-wavelength counterpart to a repeater burst has not been detected \citep[e.g.,][]{hds+17,scc+20,tvc+20a,nds+22}.

It has not yet been settled what observed behavior is necessary for FRB production (i.e., if an extreme magneto-ionic environment and/or a mechanism that provides periodic activity are a requirement or a coincidence\footnote{A case in point is the magnetar near the Galactic center, for which the observed evolution of Faraday rotation has been demonstrated to be unrelated and extrinsic to the burst production \citep{dep+18}.}) and it is unclear whether models that have been tuned to the properties of known active repeaters can be safely extrapolated to explain \emph{all} FRBs. A larger sample of repeating sources of FRBs is key to exploring similarities and differences between the population of repeaters and apparent nonrepeaters.

Optimal strategies for discovering repeating sources of FRBs are to observe FRB sky positions with more sensitive instruments or to revisit the same patch of sky repeatedly \citep{cp18}. Indeed, two FRBs discovered by the Australian Square Kilometre Array Pathfinder were shown to repeat by the more sensitive 100-m Robert C. Byrd Green Bank Telescope \citep{kso+19} and the 64-m Murriyang (Parkes) telescope \citep{ksf+21}, and FAST discovered repeat bursts from an FRB initially discovered by the Murriyang telescope \citep{lwm+20}.

The Canadian Hydrogen Mapping Experiment (CHIME) is a radio interferometer with no moving parts that observes the full Northern sky ($\delta > -11^\circ$) daily. Its FRB project \citep[CHIME/FRB;][]{abb+18} has previously reported twenty repeating sources of FRBs \citep{abb+19b,abb+19c,fab+20,bgk+21,lac+22} through a continuous survey that commenced in mid 2018.

As the number of detected FRBs continues to grow, the chance coincidence probability of finding two unrelated FRBs at similar sky position and DM becomes non-negligible. This is especially true when the sky localization precision of an FRB is coarse and the exposure towards a certain sky area is large. In this paper we present an improved and more accurate framework for the calculation of chance coincidence probabilities that builds on the approach outlined by \citet{fab+20}. We present \nsources~new repeating FRB sources discovered by CHIME/FRB from 2019 September 30 to 2021 May 1. In order to systematically search our events for repeating sources, we employed a clustering algorithm on all events in our database. We outline our source identification procedure in \S\ref{sec:obs} and source and burst characterization in \S\ref{sec:analysis}. We interpret these findings in \S\ref{sec:discussion} and conclude in \S\ref{sec:conclusions}.

Note that while preparing this manuscript, CHIME/FRB alerted the community to our discovery of new repeater FRB~20220912A, which falls outside of the cutoff date for the analysis presented here \citep{R117_atel}.  We will therefore report on it elsewhere.

\section{Observations} 
\label{sec:obs}

\subsection{CHIME/FRB}

CHIME is a radio interferometer located at $49^\circ$ latitude near Penticton, B.C. in Canada. It consists of four semi-cylindrical 100-m $\times$ 20-m reflectors with 256 dual-polarization cloverleaf antennas installed on each focal line \citep{2014SPIE.9145E..22B}. The instrument is sensitive in the range 400--800\,MHz. The FRB detection system \citep{abb+18} runs on the telescope in parallel to hydrogen mapping \citep{abb+22} and pulsar timing and search projects \citep{abb+21}, and tiles the primary beam of the telescope with four East-West $\times$ 256 North-South stationary beams that each yield 16,384 frequency channels and a 0.98304-ms time resolution. These data streams are cleaned of radio frequency interference \citep{ms22}, dedispersed with a tree algorithm using trial dispersion measures (DMs) up to approximately 13,000\,pc cm$^{-3}$ and then searched for peaks up to a width of approximately 63\,ms, all in real time. All trigger metadata are stored in a database, but only upon detection of a significant and likely astrophysical trigger are data around the trigger time saved to disk from a ring buffer. CHIME/FRB has the ability to store ``intensity'' (the beamformed Stokes-$I$ data streams) and ``baseband'' (the complex voltages measured at each of the 1,024 antennas) data. Whereas the former has a much smaller storage footprint, the latter allows for repointing the array in software \citep{mmm+21} and gives access to the native data resolution ($800 \times 10^6$ samples s$^{-1}$, in 1,024 frequency channels at 2.56 $\mu$s time resolution by default) and polarimetric information \citep{mmm+21b}, which allows for coherent removal of the dispersion delay.

\subsection{Source Identification}

Our systematic identification of new sources and associated bursts is broken into four stages. First, we obtain a list of candidate clusters by running a clustering algorithm on the coarse sky positions and DMs from detection metadata in the database (\S\ref{sec:clustering}). Then, for all bursts under consideration, we refine the measurements of their sky position and DM using the best available data (\S\ref{sec:analysis}). These measurements then form the input for a contamination rate ($R_\mathrm{cc}$) calculation, based on the probability of chance coincidence ($P_\mathrm{cc}$), that determines the likelihood of nearby events being unrelated (\S\ref{sec:pcc}). All sources with sufficiently low $R_\mathrm{cc}$ are investigated further. The $R_\mathrm{cc}$ calculation is also used to identify any outlier events in clusters (e.g., when three events were grouped together by the clustering algorithm, but only two events in the cluster are from the same source). Finally, we verify the self-consistency of all combined localizations (\S\ref{sec:localization}). All sources that pass this final test are deemed real repeaters.

Here, we have analyzed all detection metadata in the CHIME/FRB database between 2019 September 30 and 2021 May 1 with a detection signal-to-noise ratio (S/N) in the real-time search pipeline $>$ 8.5.

\subsubsection{Clustering Algorithm}
\label{sec:clustering}

We clustered events using an in-house implementation \citep{dcm+22} of the DBSCAN algorithm \citep{eksx96}. DBSCAN is an unsupervised machine learning algorithm to find clusters of points in $n$-dimensional space. We wish to find clusters in the coarse right ascension ($\alpha$), declination ($\delta$) and DM and for this, DBSCAN offers two major benefits. First, it is insensitive to the shape of the cluster as it does nearest-neighbor clustering. Bright sources in CHIME/FRB are detected in arcs of \emph{apparent} $\alpha$ and $\delta$ due to pulses/bursts being sequentially detected in the Eastern sidelobes, the main lobe and the Western sidelobes of the formed beams as the sources transit through the telescope's field-of-view.\footnote{The arcs of apparent $\alpha$ and $\delta$ are symmetric around, and peak in S/N, at the true $\alpha$ and $\delta$ of the source \citep[see Fig.~5.2 in][for an example]{ple21}.} With DBSCAN, one such arc of detections from one source is grouped in one cluster and not in multiple clusters. Secondly, DBSCAN can form an unbounded number of clusters and no arbitrary human choice of number of clusters is required.

Based on our knowledge of CHIME/FRB features and systematics, such as the sidelobe detections of bright sources, we have set the nearest-neighbor distance, $\epsilon$, to be a function of the measurement uncertainties for event parameters. The uncertainty in $\alpha$ is not constant across our declination range ($>-11^\circ$) and increases towards the North Celestial Pole (NCP; $\delta$ = 90$ ^{\circ}$): $\sigma_\alpha = 2.2\degr / \cos(\delta)$. The other tolerances are set to be constant fiducial values $\sigma_\delta = 0.5^\circ$, $\sigma_\mathrm{DM} = 13$ pc cm$^{-3}$. These are the approximate beam size full width at half maximum and the largest DM uncertainty in our tree dedispersion algorithm. The $\sigma_\mathrm{DM}$ threshold also allows for grouping of events that were detected by the real-time pipeline at suboptimal DMs (see also \S\ref{sec:completeness}). The minimum number of points in each cluster is set to be two.

This method successfully recovered all previously published CHIME/FRB-discovered repeating FRBs. After the removal of clusters from known sources\footnote{These known sources can be previously discovered repeaters or detections of known pulsars in off-meridian sidelobes, where their apparent sky positions makes them seem extragalactic from comparing their observed DMs to the maximum expected Galactic DM contributions towards their apparent sky positions.}, the analysis resulted in the identification of 65 new clusters.

\subsubsection{Chance Coincidence Probability Calculation}
\label{sec:pcc}

As the CHIME/FRB experiment continues to detect more FRBs, the probability of identifying two or more unrelated FRB sources with similar DM and within a typical localization region becomes non-negligible. To calculate the probability that each of the $x$ bursts in a cluster are physically unrelated to one another we treat the detection of FRBs as a set of independent Bernoulli trials, with each FRB detected by CHIME/FRB as one trial. For each cluster, we are then interested in the probability of detecting an additional $x - 1$ coincident bursts ($x-1$ ``successes'') that are physically unrelated (hence, independent). Coincident here means that within some volume $\Delta\alpha \Delta\delta \Delta\mathrm{DM}$ the values for burst parameters agree. The probability $P$ that another coincident burst is detected can be written according to the binomial distribution
 \begin{equation}
    P_\mathrm{cc,cluster} = \sum_{k=x-1}^{N-1} \genfrac{(}{)}{0pt}{}{N-1}{k} p_{\{\alpha, \delta, \text{DM}\}}^k (1 -  p_{\{\alpha, \delta, \text{DM}\}})^{(N-1)-k},
    \label{eq:pcc}
\end{equation}
 where $N = 2196$ is the total number of FRBs detected by CHIME/FRB\footnote{This number is subject to change as we vet FRB candidates for future catalogs. However, the $P_\mathrm{cc}$ calculation is only marginally sensitive to small variations in $N$.} by the cut-off date of this analysis and $p_{\{\alpha, \delta, \text{DM}\}}$ is the probability of a ``success'' (the probability that a source detected by CHIME/FRB is detected in the $\Delta\alpha \Delta\delta \Delta\mathrm{DM}$ volume).

The probability $p_{\{\alpha, \delta, \text{DM}\}}$ is a function of the distribution of FRB sources in the Universe and the CHIME/FRB exposure and selection function. We have estimated it empirically\footnote{With the exception of $\alpha$, for which we can assume a uniform probability distribution, as our exposure covers multiple years \citep[see also][]{fab+20}.} by using a normalized histogram of our entire FRB source sample (i.e., without repeat bursts) smoothed by a Gaussian kernel with an angular scale of 10\degr\ and a DM scale of 350\,pc\,cm$^{-3}$. Ideally, we want to consider only independent FRB sources for this analysis. It is, however, not possible to obtain a sample completely free of repeat bursts. When repeat bursts are inadvertently left in the sample, that artificially inflates $p_{\{\alpha, \delta, \text{DM}\}}$ in the respective portions of the sky, which leads to higher chance coincidence probabilities, making our analysis more conservative.

\begin{figure}
    \centering
    \includegraphics{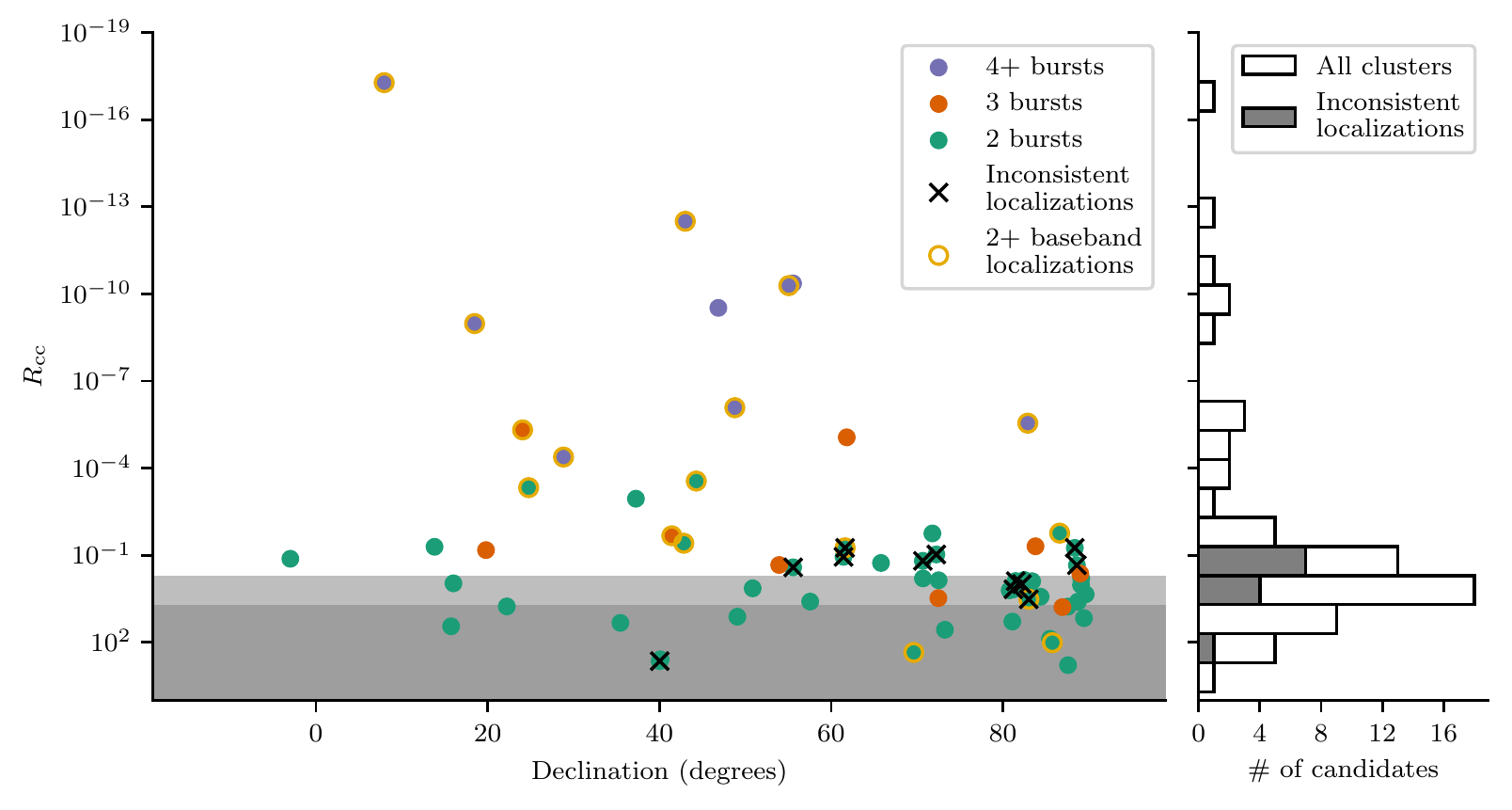}
    \caption{The panel on the left shows the minimum contamination rate ($R_\mathrm{cc}$) of all combinations of bursts in \nclusters~clusters of events as a function of the average declination of the cluster. Note that the vertical axis is inverted. The gray regions indicate $0.5 \leq R_\mathrm{cc} < 5$ (``silver'' sample; lighter region) and $R_\mathrm{cc} \geq 5$ (darker region), markers are color-coded by cluster size, and clusters with inconsistent localizations (which are then discarded) are indicated with a black cross. The panel on the right shows a histogram of the $R_\mathrm{cc}$. In general, clusters with more bursts and/or baseband localizations and clusters at lower declination have lower $R_\mathrm{cc}$.}
    \label{fig:pcc}
\end{figure}

Using $N = 2196$, as above, as our sample size/number of trials, we calculate a contamination rate (i.e., the expected number of false positives in our sample)
\begin{equation}
   R_\mathrm{cc} = P_\mathrm{cc} \times N,
\end{equation}
that is shown in Figure~\ref{fig:pcc} for each cluster. We select which sources we deem to be real repeating sources of FRBs by setting a maximum acceptable contamination rate $R_\mathrm{cc} < 0.5$ (``gold'' sample), such that there is a 50\% probability that one source in the sample got included by chance. This ``gold'' sample includes 32 previously unpublished sources. We also present clusters with $0.5 \leq R_\mathrm{cc} < 5$ (``silver'' sample) in Appendix~\ref{sec:candidates} as candidate repeaters that may warrant follow-up observations.  This ``silver'' sample includes an additional 18 previously unpublished potential repeaters. However, upon closer inspection, seven and four of those clusters, respectively, have inconsistent event localizations (see \S\ref{sec:localization} and Fig.~\ref{fig:pcc}). We thus end up with \nsources~new repeater sources (``gold'' sample) and \nappsources~new repeater candidates (``silver'' sample).

As it is possible that a cluster contains unrelated bursts (e.g., from the tolerances of the clustering analysis being large; see \S\ref{sec:clustering}), we calculate $R_\mathrm{cc}$ for each combination of events\footnote{That is, all unique combinations $\binom{n}{k}$, where $n$ is the number of events in each cluster and $2 \leq k \leq n$.} in a cluster and consider the minimum $R_\mathrm{cc}$ of all combinations (which should be the $R_\mathrm{cc}$ of all events in a cluster, if no event is an outlier). We inspect all combinations (see Fig.~\ref{fig:outliers} in Appendix~\ref{sec:extrafigs}) and flag those events that increase the $R_\mathrm{cc}$ by more than an order of magnitude as compared to the minimum $R_\mathrm{cc}$ of that cluster in one or more of the combinations (ten potential outliers in total in nine unique candidate clusters; these flagged events are marked with asterisks in Figs.~\ref{fig:gold_wfalls_0}, \ref{fig:gold_wfalls_1}--\ref{fig:gold_wfalls_3}). However, we do not exclude from our sample those flagged events, as a mix of parameter uncertainties (sky positions from real-time detection S/Ns versus baseband data and a S/N-dependent DM uncertainty) could also lead to a ``forked path'' of $R_\mathrm{cc}$ as a function of included events, where one event leads to a systematically lower $R_\mathrm{cc}$ when combined with other events in the cluster (see Fig.~\ref{fig:outliers} for examples). The outlier analysis would also reveal if two or more separate repeaters were grouped into one cluster, but we see no evidence for this. 

\section{Analysis \& Results}
\label{sec:analysis}

\begin{figure}
    \centering
    \includegraphics{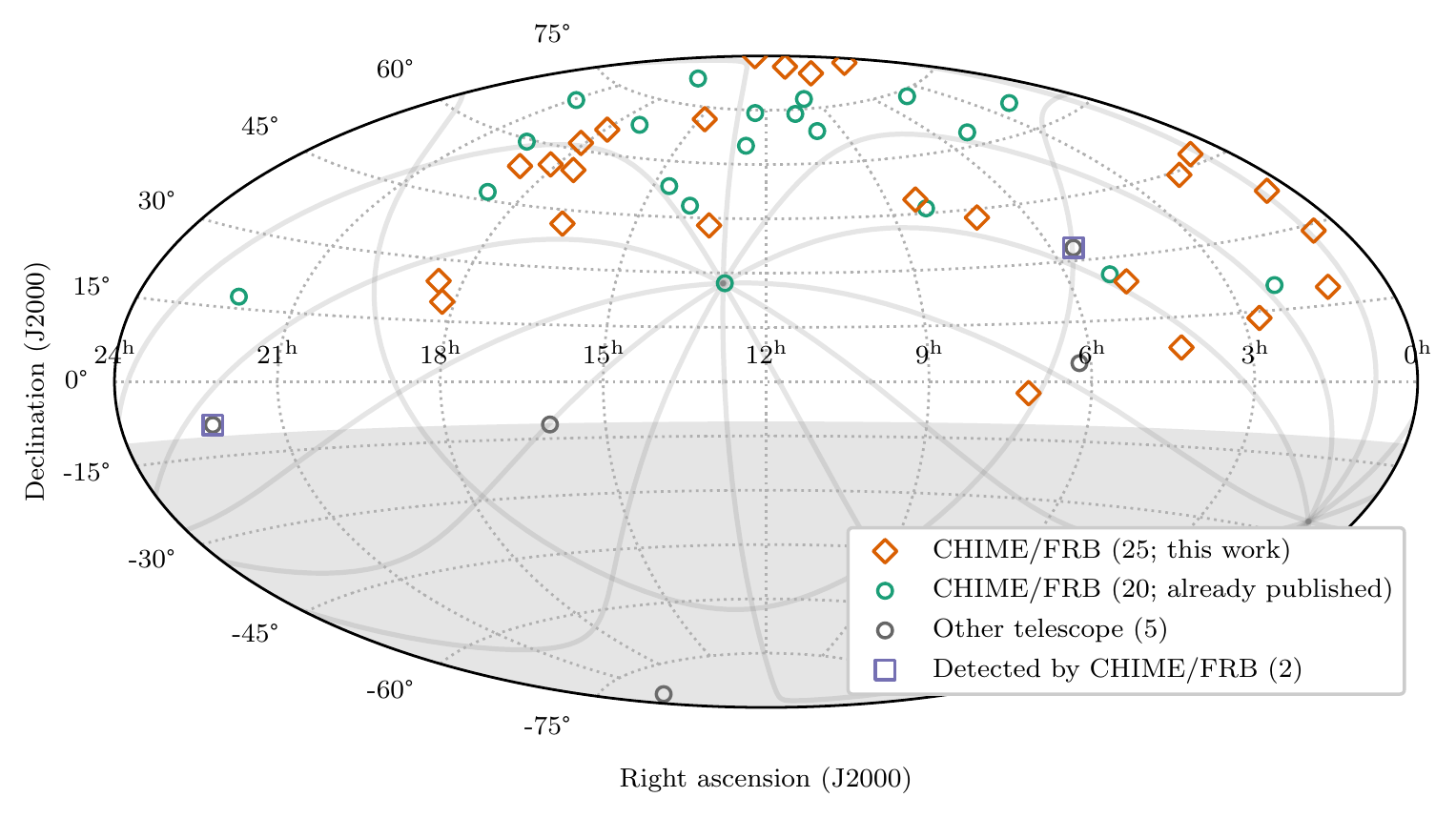}
    \caption{Sky distribution (Aitoff projection) of repeating sources of FRBs discovered by CHIME/FRB \citep[orange open diamonds and green open circles; this work and][]{abb+19b,abb+19c,fab+20,bgk+21,lac+22}, discovered by other telescopes \citep[gray open circles;][]{ssh+16a,kso+19,lwm+20,ksf+21,nal+22} and detected by CHIME/FRB \citep[purple open squares;][]{jcf+19,pat19}. $\delta < -11^\circ$ is outside of CHIME/FRB's field-of-view and shaded light gray. The gray solid lines in the background show the plane of the Galaxy ($b = 0^\circ$) as well as lines of constant Galactic longitude $0^\circ$ to $360^\circ$ in steps of $30^\circ$.}
    \label{fig:repeatersky}
\end{figure}

The properties of \nsources~new repeating sources of FRBs are summarized in Table~\ref{tab:sources} and individual burst properties are collected in a machine-readable table in the same format as the first CHIME/FRB FRB catalog \citep{aab+21}.\footnote{\url{https://www.chime-frb.ca/repeater_catalog}} A description of each field in the catalog is presented in Table~\ref{ta:catalog descriptions} in Appendix~\ref{sec:description}. A sky distribution of repeating sources of FRBs is presented in Figure~\ref{fig:repeatersky}. Dedispersed dynamic spectra of the bursts, best-fit localization regions of each source without any baseband localization (see \S\ref{sec:localization}) and the exposure up to 2021 May 1 are publicly available.\footnote{\url{https://www.canfar.net/storage/vault/list/AstroDataCitationDOI/CISTI.CANFAR/23.0004/data}; for FRBs 20200508H and 20200825B only the dynamic spectra averaged to 128 subbands are available, as the raw data for these bursts (with 16,384 frequency channels) were inadvertently deleted.}

\begin{figure}
\centering
\includegraphics[width=\textwidth]{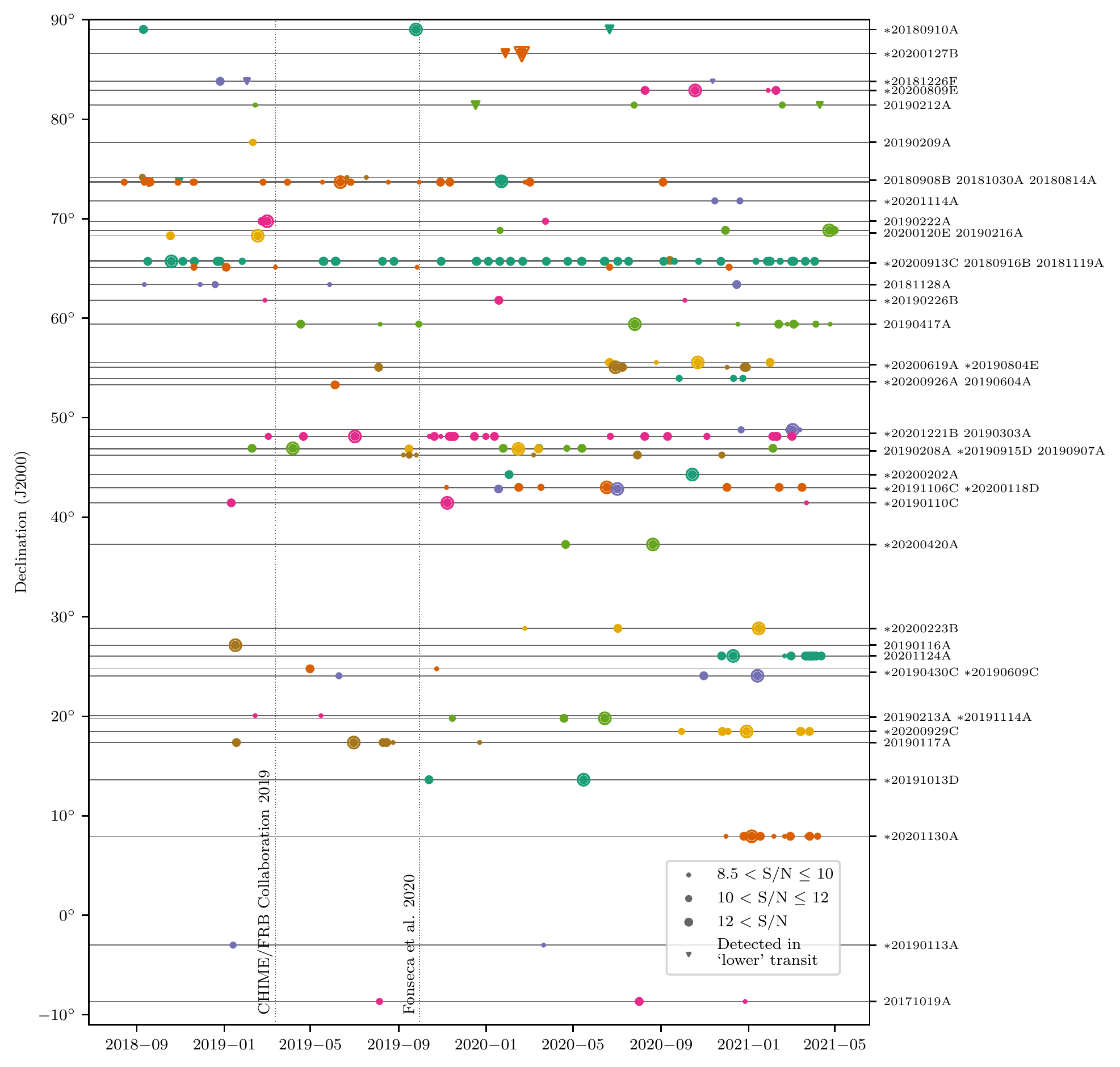}
\caption{Detection timeline for 46 repeating sources of FRBs \citep[45 discovered by CHIME/FRB and FRB~20171019A;][]{kso+19} from the start of CHIME/FRB operations to 2021 May 1. Different sources have different colors, detections in the ``lower'' transit are indicated by a downward-facing triangle and repeater discovery dates (when the second burst above a completeness threshold, here S/N $>$ 12, was detected --- not the case for all sources; see \S\ref{sec:frep}) are highlighted with an extra circle or triangle. Source names are listed on the right-hand ordinate axis, with an asterisk in front of the names of the \nsources~newly discovered repeaters. To aid readability, source names have sometimes been grouped in a row in order of descending declination. Dotted vertical lines mark the cutoff dates for \citet{abb+19c} and \citet{fab+20}.}
\label{fig:timeline}
\end{figure}

Figure~\ref{fig:timeline} shows the detection timeline of all repeating sources of FRBs discovered and detected \citep[i.e., FRB~20171019A, originally discovered by ASKAP;][]{kso+19} by CHIME/FRB up to 2021 May 1, including previously unpublished bursts from already published sources. Of the previously unpublished bursts from the known sources, we have measured their burst properties if their first detection with S/N $>$ 10 or 12 happens to be an unpublished burst for the analysis in \S\ref{sec:comparison}. We use the detection S/N and times-of-arrival of these bursts for the analyses in \S\ref{sec:reprates} and \S\ref{sec:frep}. A machine-readable table of these bursts is available in the same format and at the same location as the table with bursts from the new sources.

Results on host galaxies to the new sources will be reported by Ibik et al. (in prep.), on associated persistent radio emission by Ibik et al. (in prep.) and on cross-correlation with galaxy catalogs by Rafiei-Ravandi et al. (in prep.).

\begin{table}[t]
\begin{center}
\caption{Properties of \nsources~New Repeating Sources of FRBs, Ordered by Increasing $R_\mathrm{cc}$ (Our ``Gold'' Sample)}
\centering
\resizebox{1.05\textwidth}{!}{ 
\hspace{-1.8in}
\begin{tabular}{ccccccccccccc} \hline
    FRB Source$^a$  &  $R_\mathrm{cc}$ & $\alpha^b$ & $\delta^b$ & $l^c$ & $b^c$ & DM$^d$ & DM$_{\rm NE2001}^e$ & DM$_{\rm YMW16}^e$ & N$_{\rm bursts}$  & Exposure$^f$ & Completeness$^g$ & Burst rate$^h$ \\
          &       &  (J2000)    &  (J2000) & (deg) & (deg) & (pc~cm$^{-3}$) & (pc~cm$^{-3}$) &(pc~cm$^{-3}$) &      &     (hr, upper / lower) &  (Jy ms) & (hr$^{-1}$) \\ \hline
20201130A & $5.3 \times 10^{-18}$ & 64.39(1) & 7.94(1) & $185.4$ & $-29.0$ & 287.984(7) & 56 & 69 & 12 & $15 \pm 11$ & $10.4$ & 2.34$_{-2.01}^{+2.28}$ \\
20191106C$^i$ & $3.1 \times 10^{-13}$ & 199.58(1) & 43.002(9) & $105.7$ & $73.2$ & 333.40(2) & 25 & 21 & 7 & $66 \pm 3$ & $1.9$ & $(2.54_{-1.35}^{+2.23}) \times 10^{-2}$ \\
20200619A & $4.3 \times 10^{-11}$ & 272.6(1) & 55.56(6) & $84.0$ & $27.8$ & 439.772(4) & 51 & 45 & 5 & $110 \pm 6$ & $2.5$ & $(1.3_{-0.9}^{+1.7}) \times 10^{-2}$ \\
20190804E & $5.2 \times 10^{-11}$ & 261.34(2) & 55.069(8) & $82.9$ & $34.1$ & 363.68(1) & 43 & 37 & 6 & $94 \pm 4$ & $1.9$ & $(1.52_{-0.86}^{+1.49}) \times 10^{-2}$ \\
20190915D & $3 \times 10^{-10}$ & 11.78(3) & 46.86(2) & $122.2$ & $-16.0$ & 488.69(2) & 89 & 88 & 7 & $92 \pm 1$ & $11.3$ & $(2.56_{-1.36}^{+2.25}) \times 10^{-1}$ \\
20200929C$^i$ & $10^{-9}$ & 17.04(2) & 18.47(1) & $128.5$ & $-44.2$ & 413.659(4) & 38 & 29 & 7 & $56 \pm 2$ & $5.5$ & $(1.45_{-0.77}^{+1.27}) \times 10^{-1}$ \\
20201221B$^i$ & $8.3 \times 10^{-7}$ & 124.20(3) & 48.78(2) & $170.5$ & $33.8$ & 510.42(5) & 51 & 46 & 6 & $90 \pm 2$ & $3.4$ & $(3.78_{-2.13}^{+3.68}) \times 10^{-2}$ \\
20200809E & $2.8 \times 10^{-6}$ & 20.0(1) & 82.89(2) & $123.9$ & $20.1$ & 1703.48(1) & 74 & 83 & 4 & $510 \pm 2$ / $410 \pm 26$ & 2.5 / 6.1 & $(2.81_{-1.85}^{+3.61}) \times 10^{-3}$ \\
20190609C & $4.8 \times 10^{-6}$ & 73.32(1) & 24.068(6) & $177.4$ & $-12.4$ & 479.602(9) & 113 & 154 & 3 & $72 \pm 0$ & $2.1$ & $(7.3_{-6.0}^{+15.7}) \times 10^{-3}$ \\
20190226B$^j$ & $8.7 \times 10^{-6}$ & $273.62_{-0.34}^{+0.17}$ & $61.67_{-0.14}^{+0.41}$ & $90.9$ & $28.0$ & 632.597(9) & 51 & 45 & 3 & $85 \pm 52$ & $1.1$ & $(3.65_{-3.47}^{+6.20})\times 10^{-3}$  \\
 & & $270.25_{-0.24}^{+0.38}$ & $61.93_{-0.41}^{+0.15}$ & $91.1$ & $29.6$ & & 48 & 43 & & & \\
20200223B$^i$ & $4.1 \times 10^{-5}$ & 8.265(8) & 28.831(7) & $118.1$ & $-33.9$ & 202.268(7) & 46 & 37 & 5 & $52 \pm 2$ & $7.6$ & $(1.44_{-0.95}^{+1.86}) \times 10^{-1}$  \\
20200202A & $0.00028$ & 25.93(2) & 44.290(7) & $132.7$ & $-17.6$ & 722.37(7) & 83 & 84 & 2 & $62 \pm 5$ & $6.1$ & $(6.53_{-4.78}^{+10.40})\times 10^{-2}$ \\
20190430C & $0.00047$ & 277.210(9) & 24.770(9) & $53.1$ & $15.7$ & 400.57(1) & 100 & 84 & 2 & $29 \pm 7$ & $11.5$ & $(2.36_{-2.02}^{+5.11}) \times 10^{-1}$ \\
20200420A & $0.0011$ & $7.57_{-0.02}^{+0.04}$ & $37.26_{-0.06}^{+0.04}$ & $118.3$ & $-25.4$ & 671.5(4) & 58 & 50 & 2 & $55 \pm 29$ & $0.7$ & $(9.58_{-10.40}^{+36.20})\times 10^{-4}$ \\
20200127B & $0.017$ & 119.2(1) & 86.609(8) & $126.6$ & $28.0$ & 351.3(4) & 54 & 51 & 2 & $154 \pm 13$ / $951 \pm 18$ & 3.7 / 5.3 & $< 1.05 \times 10^{-2}$\\
20201114A & $0.018$ & 221.59(8) & 71.79(3) & $111.2$ & $42.6$ & 322.23(1) & 38 & 31 & 2 & $135 \pm 27$ / $83 \pm 17$ & 6.8 / 66.3 &$(2.33_{-1.98}^{+5.03})\times 10^{-2}$ \\
20190110C & $0.021$ & 249.33(1) & 41.445(9) & $65.6$ & $42.1$ & 221.92(1) & 37 & 30 & 3 & $45 \pm 11$ & $1.2$ & $(8_{-6}^{+13})\times10^{-3}$ \\
20200118D & $0.038$ & 106.91(1) & 42.837(9) & $174.4$ & $20.9$ & 625.57(3) & 77 & 92 & 2 & $72 \pm 5$ & $1.2$ & $(3.3_{-2.7}^{+7.1}) \times 10^{-3}$ \\
20181226F & $0.049$ & $95.87_{-0.31}^{+0.23}$ & $83.80_{-0.16}^{+0.47}$ & $129.8$ & $26.2$ & 241.887(3) & 58 & 56 & 3 & $394 \pm 192$ / $376 \pm 215$ & 4.8 / 10.6 & $(2.37_{-2.53}^{+8.96}) \times 10^{-3}$ \\
20191013D & $0.051$ & 40.42(2) & 13.63(3) & $159.7$ & $-41.3$ & 523.57(3) & 43 & 36 & 2 & $67 \pm 1$ & $8.9$ & $(7.09_{-5.83}^{+15.20}) \times 10^{-2}$ \\
20191114A & $0.066$ & 273.57(3) & 19.80(4) & $47.0$ & $16.8$ & 552.47(6) & 98 & 80 & 3 & $61 \pm 5$ & $2.1$ & $(1.36_{-0.99}^{+2.16})\times 10^{-2}$ \\
20190113A$^j$ & $0.13$ & $108.26_{-0.26}^{+0.29}$ & $-2.98_{-0.52}^{+0.51}$ & $218.1$ & $3.4$ & 428.90(1) & 178 & 251 & 2 & $40 \pm 25$ & $5.0$ & $(4.97_{-5.13}^{+11.10})\times 10^{-2}$ \\
 & & $106.56_{-0.80}^{+0.78}$ & $-3.01_{-0.56}^{+0.54}$ & $217.3$ & $1.9$ & & 187 & 275 & & & \\
20200913C & $0.18$ & 269.03(6) & 65.80(3) & $95.5$ & $30.2$ & 576.880(3) & 48 & 43 & 2 & $156 \pm 3$ & $1.8$ & $(2.85_{-2.34}^{+6.12})\times 10^{-3}$ \\
20200926A & $0.22$ & $283.28_{-0.20}^{+0.18}$ & $53.95_{-0.31}^{+0.30}$ & $83.7$ & $21.4$ & 758.44(5) & 64 & 59 & 3 & $67 \pm 43$ & $0.5$ & $(1.58_{-1.53}^{+2.70})\times10^{-3}$ \\
20180910A & $0.43$ & 354.8(9) & 89.01(1) & $122.6$ & $26.2$ & 684.408(1) & 56 & 55 & 3 & $3913 \pm 270$ / $4335 \pm 26$ & 19.1 / 30.7 & $(3.83_{-3.16}^{+8.23}) \times 10^{-3}$ \\
\hline
\end{tabular}
}
\label{tab:sources}
\end{center}
$^a$ Here we employ the TNS naming convention, see \url{https://www.wis-tns.org/astronotes/astronote/2020-70}. \\
$^b$ Positions were determined from baseband data (symmetric uncertainties) where available and from per-burst S/N data otherwise (asymmetric uncertainties; see \S\ref{sec:localization}), and uncertainties are at 90\% confidence. \\
$^c$ Galactic longitude and latitude for the best position. \\
$^d$ Inverse-variance-weighted average DM, and the uncertainty in the calculation of that average. In cases where the DM was fixed in the \texttt{fitburst} fit, we set the DM uncertainty to 0.5\,pc\,cm$^{-3}$ for the calculation of the average. For candidates with large DM variation from burst to burst, this average and its uncertainty may not necessarily be representative and we advise looking at the individual burst DMs using generous DM ranges when conducting follow-up searches of these candidates. \\
$^e$ Maximum model prediction along this line-of-sight for the NE2001 \citep{ne2001} and YMW16 \citep{ymw17} Galactic electron density distribution models. Neither model accounts for DM contributions from the Galactic halo, which contributes no more than 111\,pc\,cm$^{-3}$ \citep{cbg+23}. \\
$^f$ For sources observed twice a day, the second entry corresponds to the less sensitive lower transit. The uncertainties in the total exposure for the upper and lower transits of each source are dominated by the corresponding source declination uncertainties since the widths of the synthesized beams vary significantly with declination. \\
$^g$ Fluence completeness limits are given at the 90\% confidence level. For sources observed twice a day, the second entry corresponds to the less sensitive lower transit. \\
$^h$ Adjusted to a 5\,Jy\,ms fluence threshold, assuming a $-1.5$ power-law energy index. \\
$^i$ The best-known sky position of this source falls in between the FWHMs at 600\,MHz of the synthesized beams. We report the exposure at the beam centre of the nearest beam and have scaled the fluence completeness threshold accordingly. \\
$^j$ Two equally likely ``islands'' in the localization 90\% confidence region.
\end{table}
\normalsize

\subsection{Localization}
\label{sec:localization}

For the purposes of clustering events to obtain repeater candidates, we use the positions from the real-time detection pipeline (``header'' localization), which are based solely on the per-beam S/N values. This localization method has been used for past CHIME/FRB publications; most recently \citet{aab+21}.  For an in-depth description of this localization methodology, see \citet{abb+19c}. Briefly, the method is a grid-search $\chi^2$ minimization, where observed ratios among per-beam S/N values are compared to predictions from the CHIME/FRB beam model\footnote{\url{https://github.com/chime-frb-open-data/chime-frb-beam-model}}, over a grid of possible sky locations. The method can be applied to multiple bursts from the same source by summing the $\chi^2$ values for each data set. As the best-fit sky position, we construct a 90\% confidence region. For most sources, this forms one ``island'', but for a few sources (FRB~20190226B and FRB~20190113A) two islands are equally likely, and we present two best-fit localizations in Table~\ref{tab:sources}.

For those bursts for which baseband data were saved to disk, we run the baseband pipeline detailed by \citet{mmm+21} to generate a sky-localization. This pipeline works in two main stages. First, a crude grid of tied-array beams is formed to determine whether the burst occurred in a sidelobe, and to obtain an initial position estimate. Next, a tightly-packed grid of beams is formed around the initial position to get a refined best-estimate position. In $\sim$20\% of cases, the baseband pipeline does not converge, due to a combination of low S/N, dedispersion artefacts and other reasons. Work to debug these issues is underway, and we defer those baseband localizations to future work. We do still include the events in our analysis, but use their positions from the real-time detection pipeline.

When one or more baseband localizations are available, a combined position is calculated using the inverse-variance weighted mean. The uncertainties on the combined sky positions are likely overestimated, as the systematic uncertainties are accounted for in each individual error bar \citep{mmm+21}. We defer a more precise treatment of the combined systematics to future work. When no baseband localization for a source is available, a final best sky position for each source is calculated from combining header localizations as was done by \citet{fab+20}.\footnote{The only difference is that here, we do not attempt to rule out regions of parameter space by comparing the spectral occupancy of the bursts with the spectral sensitivity predicted by our beam model.}

When labelling a collection of events as a candidate repeater, localization consistency is crucial to check. The nature of this check depends on the available data. For candidates with multiple baseband triggers, we verify the positions are consistent such that a weighted average yields no outliers. We also ensure any given baseband position is consistent with the uncertainty region derived from the header data alone, and we verify that each position is localized to the correct lobe of the formed beams and not a sidelobe. Baseband positions that are contained within the 90\% confidence region are considered fully consistent, those in close proximity (within a few arcminutes) are considered marginal, and those several arcminutes to a few degrees away are considered outliers.  For all repeater candidates, a successful joint-fit header localization is the baseline requirement. This test fails for seven clusters in the ``gold'' and four clusters in the ``silver'' sample; see \S\ref{sec:pcc} and Fig.~\ref{fig:pcc}, and those clusters are discarded. Conceptually, the joint fit is similar to taking the intersection of the uncertainty regions of individual bursts. Individual uncertainty regions reflect the instantaneous sensitivity of the beams that resulted in the detection, as well as those that did not. A merged localization with \mbox{(near-)zero} sky area is indicative of bursts that do not originate from the same sky location. We take all combinations of events from a cluster and check the merged localizations. If just one event led to the \mbox{(near-)zero} sky area, we discard that event. If all events in a cluster are inconsistent with each other, we discard the full cluster.

\begin{figure}
\centering
\includegraphics{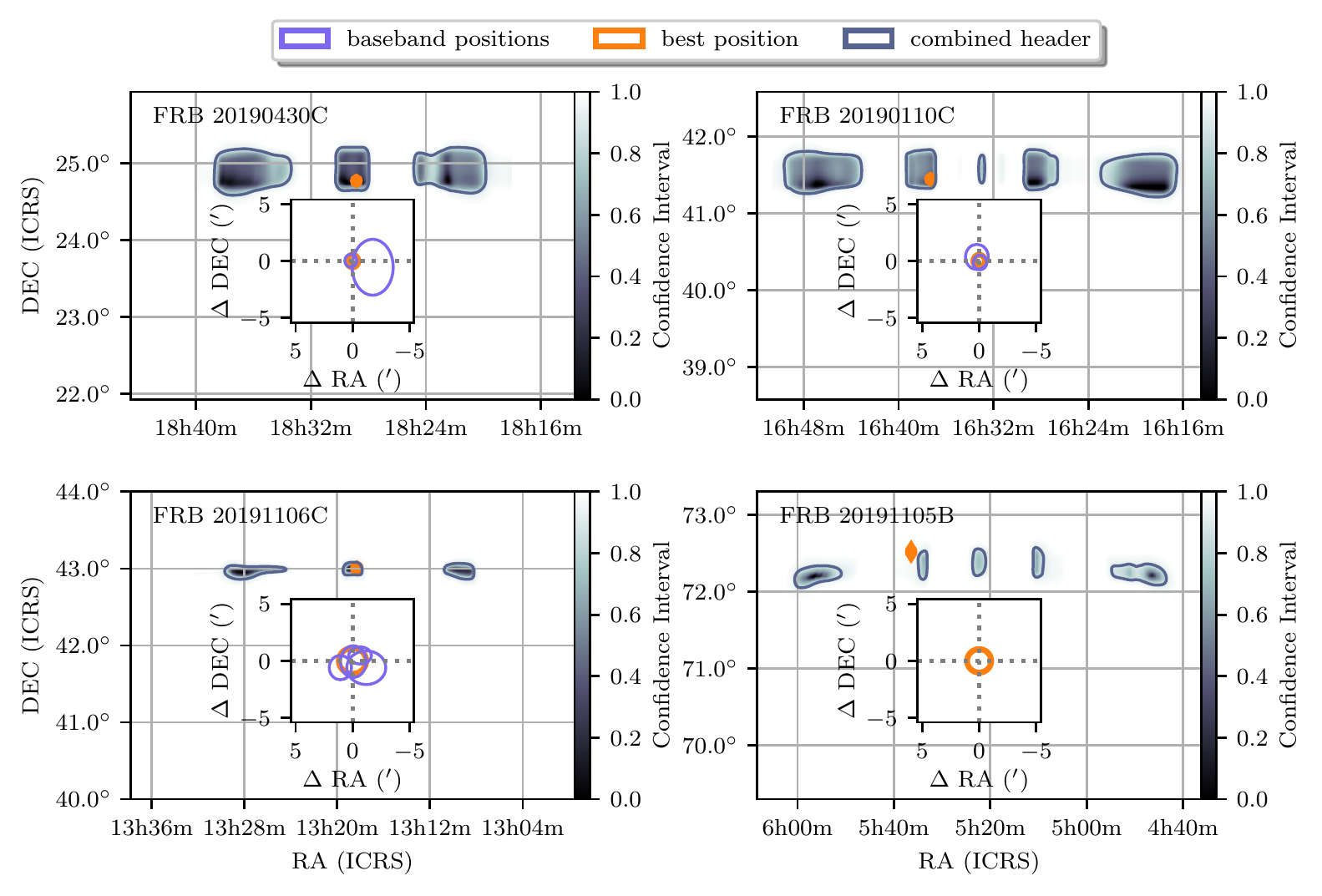}
\caption{Four representative examples of combined localization patterns, as a function of right ascension (RA) and declination (DEC). The gray shaded area shows the 90\% confidence region of the combined header (real-time detection) localization with the best-known position overlaid in orange. The insets show a zoom-in around the combined baseband localization (orange) on top of the individual baseband localizations (purple) shown as well. The combined header localizations allow for multiple possible degenerate islands and the baseband localizations can break that degeneracy.}
\label{fig:localization}
\end{figure}

Figure~\ref{fig:localization} shows four source localizations that represent four possible ways in which the combined header and (combined) baseband positions can converge to the true sky position of a source. In the example of FRB~20190430C (top left), multiple header localizations do not greatly reduce the allowed phase space, and the baseband positions land on the main island. In the example of FRB~20190110C (top right), the central region is ruled out by combining multiple header localizations, and this is confirmed by the positions derived from baseband data. In the examples of FRBs~20191106C (bottom left) and 20191105B (bottom right), combining multiple header localizations reduces the allowed phase space in the North-South/declination direction. The baseband positions fall on the mainlobe island for the former source, but on a sidelobe island for the latter source.

\newpage

\subsection{Burst and Source Properties}
\label{sec:props}

We characterize each new burst as in the first CHIME/FRB catalog \citep{aab+21}. In summary, we employ a least-squares fitting algorithm that allows for multiple components and provides measurements of DM and scattering (for each burst) and times-of-arrival, widths and spectral properties (for each burst component). Width and scattering timescales can be robustly measured from CHIME/FRB intensity data for values larger than $100\,\mu$s. Additionally, we determine a bandwidth and duration from the burst dynamic spectrum dedispersed to the best-fit DM. Flux and fluence are determined from converting the measured intensities to physical quantities using daily observations of bright calibration sources with known flux densities. CHIME/FRB exposure and sensitivity are calculated based on metrics collected that included instrument uptime and monitoring of detection S/Ns of known pulsars. Finally, we determine a fluence detection threshold for each source to characterize our completeness. This is based on a Monte Carlo analysis that takes into account uncertainties in the sky position and the spectral properties of the source. Burst rates are calculated from the number of detections and exposure to a source. While approximately 19 FRBs in the sample show downward-drifting subbursts by eye, obtaining robust drift rate measurements is challenging due to low detection S/N. We defer such measurements to later work.

\begin{figure}
\centering
\includegraphics[width=0.9\textwidth]{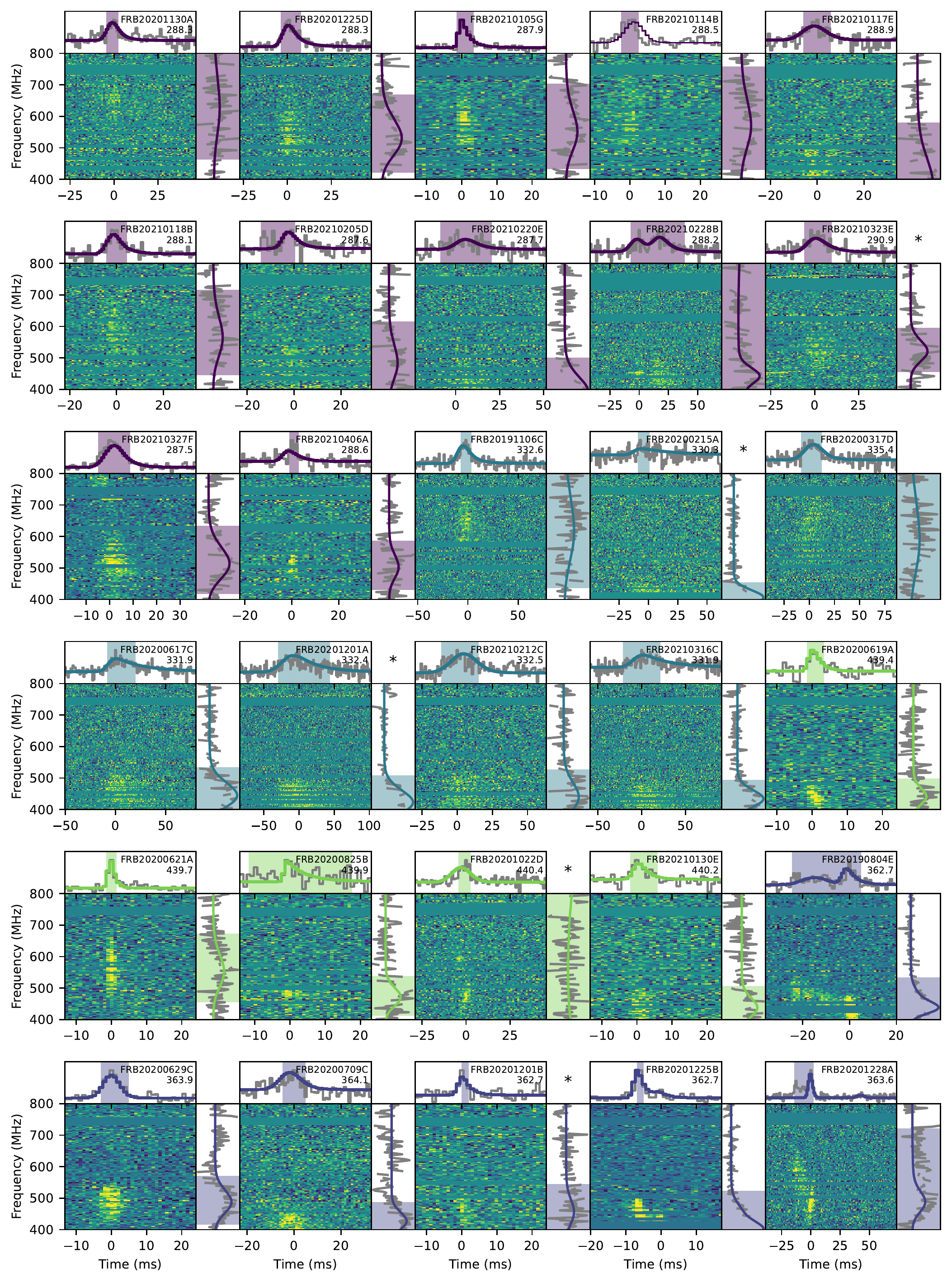}
\caption{Dynamic spectra (``waterfall plots''), frequency-averaged time series and time-averaged spectra for all sources, ordered from high to low $R_\mathrm{cc}$. Bursts from the same source are grouped by the same color. In the upper right-hand corner of each panel are the the TNS names and DM values in units of pc\,cm$^{-3}$. The model fits for which scattering was determined to be insignificant are over-plotted in the time series and spectra with \textit{thin} blue lines, whereas the fits for which scattering was determined to be significant are over-plotted in the time series and spectra with \textit{thick} blue lines. The colored shaded regions in the time series and frequency spectra span the respective full width at tenth of maximum (FWTM) burst widths and emission bandwidths. An asterisk in the top right corner of a panel indicates that the event is a potential outlier, based on how it changes the $R_\mathrm{cc}$ of the cluster it belongs to (see \S\ref{sec:pcc}). Continued in Figs.~\ref{fig:gold_wfalls_1}--\ref{fig:gold_wfalls_3}.}
\label{fig:gold_wfalls_0}
\end{figure}

Figures~\ref{fig:gold_wfalls_0}, \ref{fig:gold_wfalls_1}--\ref{fig:gold_wfalls_3} show dynamic spectra for each burst of each source. As before, most repeater bursts show narrow bandwidths of typically 50--200\,MHz, with some sources showing bursts with clear cases of downward-drifting subbursts (FRBs~20190804E, 20190915D, 20200929C, 20200809E, 20201014B, 20200420A, 20201114A and 20200118D) of a few to tens of MHz\,ms$^{-1}$. Interestingly, FRBs~20200127B and 20181226F at modest extragalactic DMs ($\lesssim 300$\,pc\,cm$^{-3}$) show larger bandwidths and narrower widths than is commonly observed for repeaters, reminiscent of bursts from other nearby sources FRBs~20181030A \citep{bkm+21} and 20200120E \citep{bgk+21}. Some notable bursts: FRB~20210327F of source FRB~20201130A shows two resolved components spaced by about 300 MHz and a few seconds, FRBs~20200214A, 20200313C and 20200313B of source FRB~20190915D show two subbursts separated by $\sim$60, 30 and 30\,ms, respectively, and FRB~20201014B of source FRB~20200202A shows at least seven subbursts with a spacing that shows a hint of possible periodicity, which is not statistically significant.

A preliminary long-term periodicity search with an algorithm that uses a Pearson's $\chi^2$ test \citep[e.g.,][]{aab+20} for all sources yielded no detections, which is not surprising as the largest number of bursts detected for a single source in this sample is only 12.

\subsection{Comparison Analysis}
\label{sec:comparison}

Whether repeating FRBs represent a distinct population of astrophysical objects from those that have not been seen to repeat remains an open question. It is possible that, in time, all FRB sources will eventually emit repeat bursts. However, wait times between bursts may be long. It is therefore worthwhile to investigate other possible differences between the repeating and apparently-nonrepeating classes of FRBs.

For this reason, \citet{aab+21} reported on a statistical comparison of repeater and apparent nonrepeater burst properties.  This study revealed significant differences in both burst width and bandwidth, as studied in much greater detail by \citet{pgk+21}. However, no other statistically significant differences in morphological properties (e.g., DM, scattering) were identified. 

Here we repeat the statistical comparison first reported in \citet{aab+21}, now including the repeater sample reported on here, together with all other published CHIME/FRB-discovered repeaters \citep{abb+19b,abb+19c,fab+20,bgk+21,lac+22,kso+19}. In this analysis, we compare all the same properties as for \citet{aab+21}, between the new, much larger sample of repeating FRB sources, and the Catalog 1 apparent nonrepeaters. We also compare the previously published CHIME/FRB repeater sample with the new repeater sample, to compare samples discovered with different total sky exposures. The more detailed study of \citet{pgk+21} is beyond our scope, and also unlikely very informative as the differences in morphology and spectra clearly persist strongly (see \S\ref{sec:rep_vs_nonrep} below).

As in the \citet{aab+21} analysis, we apply the following constraints when defining our comparison samples:
\begin{itemize}
    \item Events with \texttt{bonsai} (real-time detection pipeline) S/N $<$ 12 are rejected to mitigate completeness issues.
    \item Events with DM $<$ 1.5 max($\textrm{DM}_{\textrm{NE2001}}$, $\textrm{DM}_{\textrm{YMW16}}$) are rejected to avoid misidentifying rotating radio transients or radio pulsars as FRBs due to $\sim$50\% errors from models used to estimate the Galactic DM.  
    \item Events detected in sidelobes are rejected\footnote{Detections in far sidelobes show multiple narrow knots of emission ($\sim$10-MHz wide each, separated by tens of MHz) due to the chromatic beam response in the far sidelobes (Lin et al. in prep.).} since our understanding of the shape of the primary beam at large zenith angles is poor, rendering event characterization challenging.
\end{itemize}
With these cuts, 155 events are rejected from the nonrepeaters sample, four sources are rejected from the previously published CHIME/FRB repeaters sample, and 18 sources are rejected from our new repeaters sample. The remaining events represent a total of 305 apparent nonrepeaters, 14 previously published CHIME/FRB repeaters, 26 new repeaters (22 from the repeater sample reported here and four\footnote{FRBs~20190907A \citep{fab+20}, 20200120E \citep{bgk+21}, 20171019A \citep[originally discovered by ASKAP;][]{kso+19} and 20201124A \citep{lac+22}.} from other CHIME/FRB-discovered repeaters not included in Catalog 1), for a total of 40 CHIME/FRB repeating sources.\footnote{In the Catalog 1 sample, two repeater bursts and four nonrepeaters do not have their flux and fluence measured, as these bursts occured right after a system restart with no calibration available and the comparison samples for flux and fluence are thus smaller by six.}

Note that below, we also relax some of these constraints in order to probe sensitivity of source selection in our comparison analysis. One significant difference in this analysis compared to that reported by \citet{aab+21} is that we no longer require the first-detected burst from a repeater to exceed the S/N cutoff, but rather that {\it any} of the detected bursts exceed the S/N cutoff. This is because our repeaters are now identified through a clustering algorithm that can find repeating sources whose first event is not very bright (i.e., we consider all events with S/N $>$ 8.5). In our analysis described next, we use the first repeating event from each repeating source discovered by CHIME/FRB that satisfies all the conditions listed above.

For distribution comparisons with uncensored data (i.e., those without upper or lower limits), we report both Anderson–Darling (AD) and Kolmogorov–Smirnov (KS) tests. A $p$-value of $<$ 0.01 indicates 99\% confidence that two samples originate from different distributions. For censored data, we apply statistical survival tests that consider upper and lower limits because we lose information if we directly omit censored data. For properties that have upper limits (pulse width and scattering time), we use both the Peto \& Peto test and the log-rank test, using R NADA package’s \texttt{cendiff} function\footnote{\url{https://rdrr.io/cran/NADA/man/cendiff.html}}  \citep{helsel2005nondetects, NADApackage, Rsoftware}. For events with properties containing lower limits (fluence and flux), a right censored survival test is applied using R NADA package's \texttt{survdiff} function\footnote{\url{https://rdrr.io/cran/survival/man/survdiff.html}} \citep{NADApackage,Rsoftware,HarringtonFleming1982}.

\subsubsection{New versus Previously Published CHIME/FRB Repeater Comparison}
\label{sec:old_vs_new}

Here we compare properties of previously published and new CHIME/FRB repeaters to determine whether the sample discovered with longer total exposure time differs from that discovered sooner. For example, a sample discovered with longer exposure time might be expected to have a larger mean DM, since a longer exposure may provide more distant sources sufficient time to produce two detectable bursts.

We compare distributions of right ascension, declination, DM, eDM, \texttt{bonsai} S/N, \texttt{fitburst} S/N, boxcar width, bandwidth, fluence, flux, width, and scattering. Here, eDM is defined as the DM after subtracting the NE2001 expected maximum Galactic contribution towards the source. In all cases except flux\footnote{Log-rank test $p$-value = 0.00357 and Peto $\&$ Peto test $p$-value = 0.0102.}, $p$-values are $>$ 0.01. However, we note that lower limits dominate in the fluxes in both the published sample and the new sample (all fluxes in the published sample are censored). Additionally, the fluxes of the new repeaters are more reliable due to better localizations while the fluxes of the previously published repeaters have more uncertainty. Thus, we cannot be confident of the difference in flux. We conclude that there is a lack of significant differences between the two samples’ properties. This may be due to insufficient difference in exposure time for samples of these sizes to yield a significant detection, or too few sources to permit any difference detection.

\subsubsection{Repeater versus Apparent Nonrepeater Comparison}
\label{sec:rep_vs_nonrep}

The $p$-value results of our statistical tests comparing repeaters and apparent nonrepeaters are shown in Table~\ref{ta:rep_vs_nonrep}. For distributions of right ascension, declination, \texttt{bonsai} S/N, \texttt{fitburst} S/N, fluence, and flux, we find that the distributions for repeaters and nonrepeaters are consistent with having been drawn from the same population, as we did in our prior analysis. For burst width (both measured and box car) and spectral parameters, we continue to see a statistically significant distinction between distributions, consistent with that previously studied by \citet{pgk+21}, which will be further quantified in a future paper. \cite{pgk+21} found $p$-value = $10^{-3}$ when comparing width between repeaters and nonrepeaters, and here, we find $p_{AD} = 1.88\times 10^{-8}$ and $p_{KS} = 3.73 \times 10^{-5}$. For bandwidth, \cite{pgk+21} observed $p$-value = $10^{-5}$, and we observe $p_{AD} = 1.24\times 10^{-8}$ and $p_{KS} = 5.35 \times 10^{-8}$. \cite{pgk+21} obtained $p$-value = $10^{-4}$ for boxcar width, and we get $p_{AD} = 2.49\times 10^{-5}$ and $p_{KS} = 6.20 \times 10^{-5}$ in our analysis. Furthermore, with the new, larger sample of repeaters, we now detect a new potential difference in the two distributions, as detailed next.

\begin{figure}
\centering
\includegraphics{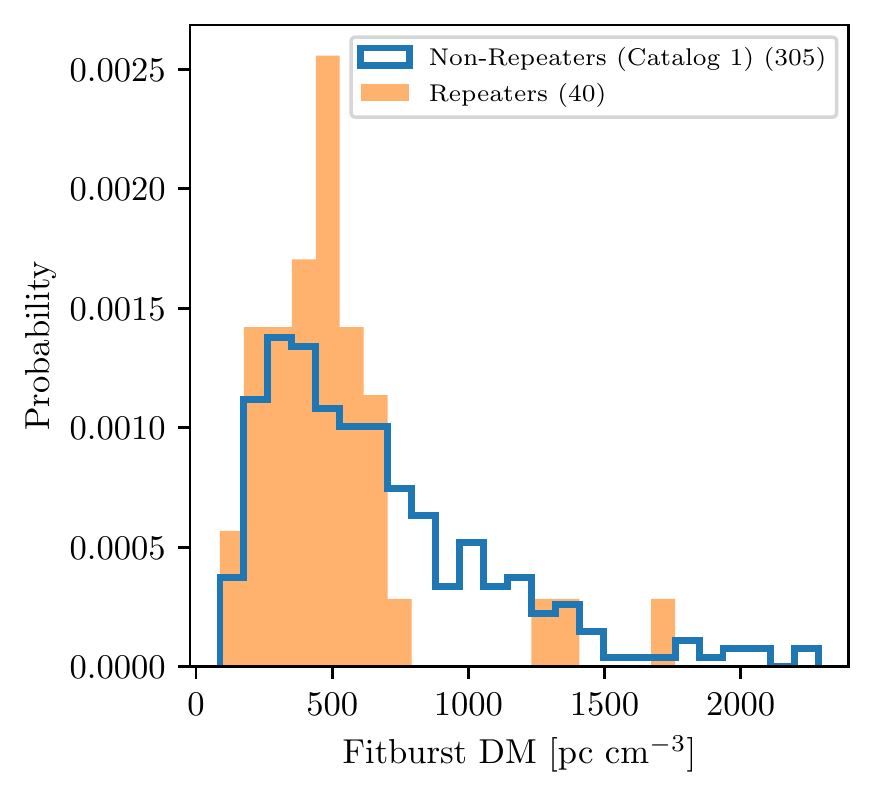}
\includegraphics{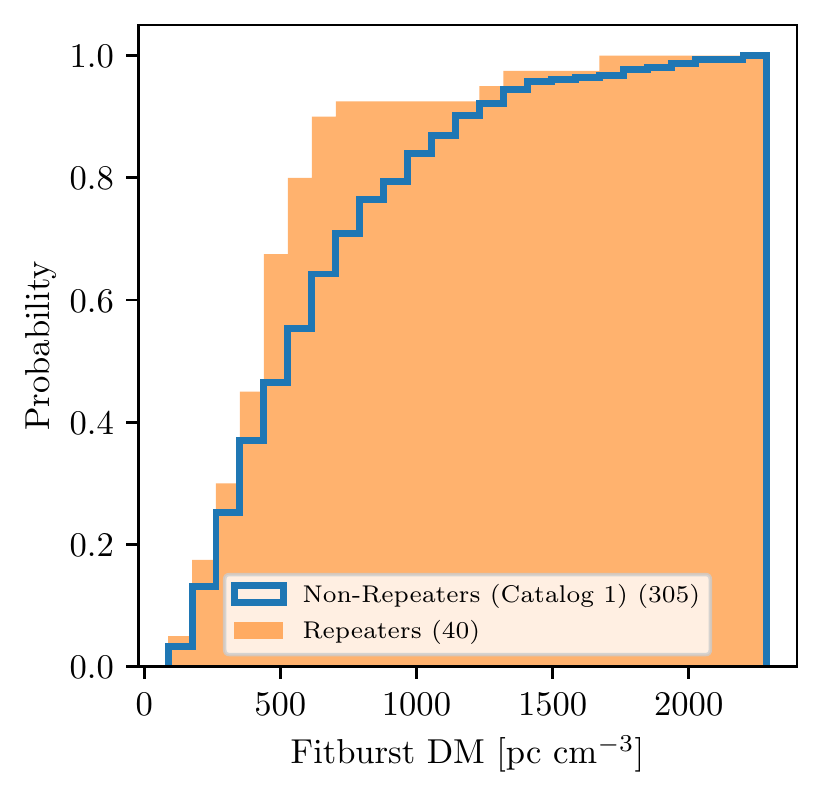}
\caption{Normalized histogram (left) and cumulative distribution (right) of the DM of the 305 nonrepeating events from \citet[][blue]{aab+21} and the complete sample of 40 CHIME/FRB-discovered repeating sources of FRBs (orange).}
\label{fig:dm_oneoffvsrepeater}
\end{figure}

\begin{figure}
\centering
\includegraphics{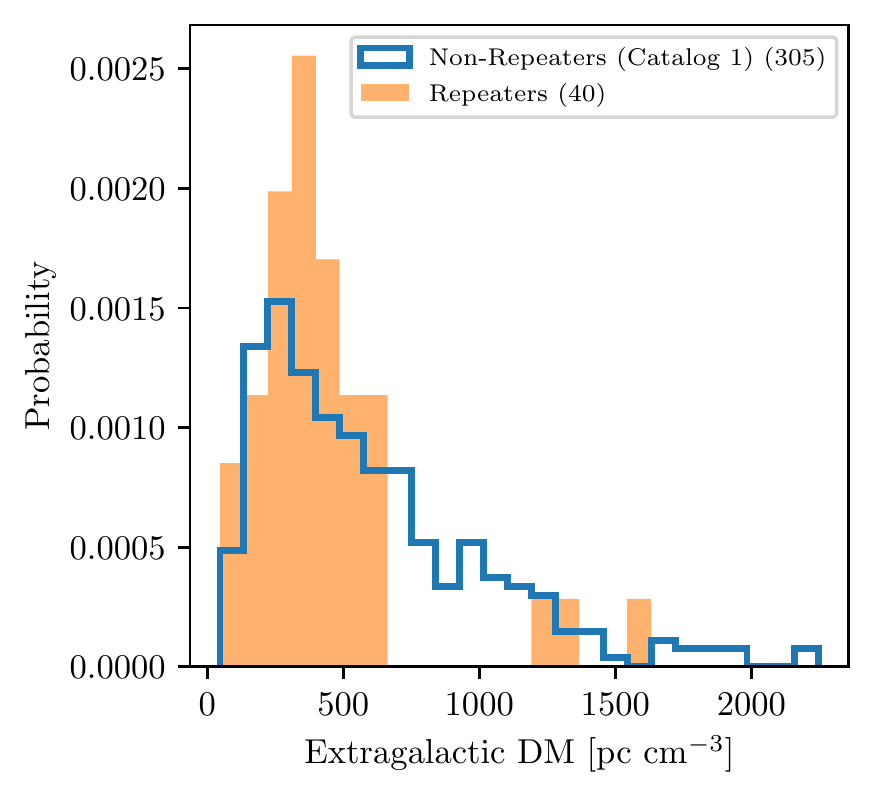}
\includegraphics{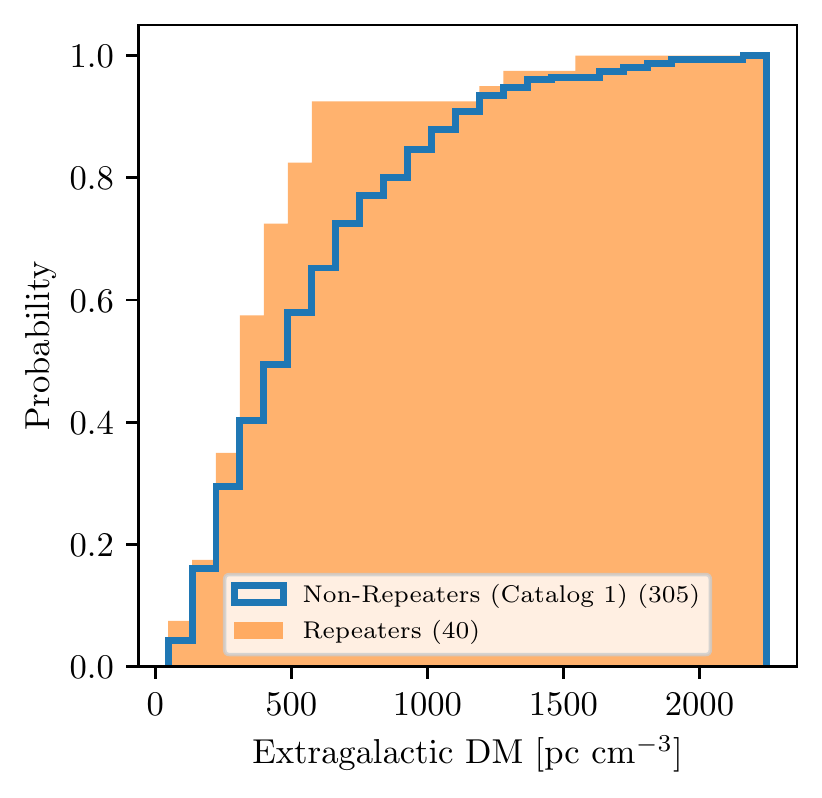}
\caption{Normalized histogram (left) and cumulative distribution (right) of the eDM of the 305 nonrepeating events from \citet[][blue]{aab+21} and the complete sample of 40 CHIME/FRB-discovered repeating sources of FRBs (orange).}
\label{fig:edm_oneoffvsrepeater}
\end{figure}

\begin{figure}
\centering
\includegraphics[width=\textwidth]{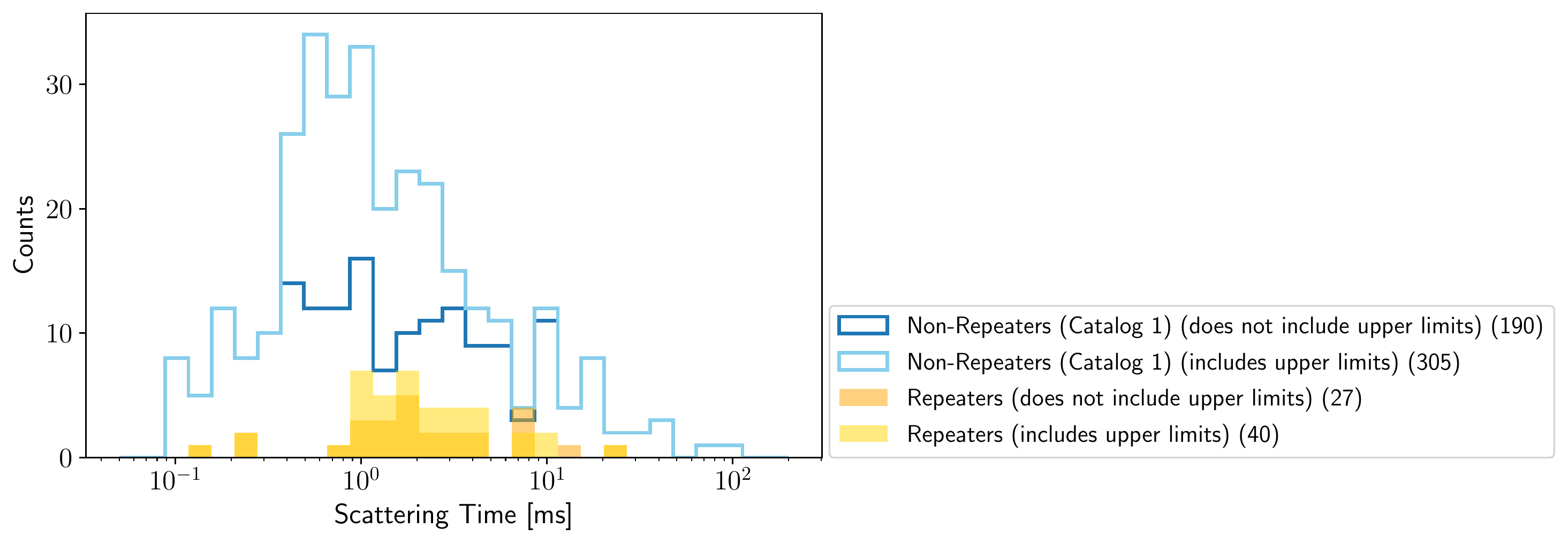}
\caption{Histogram  of the scattering times of the 305 nonrepeating events from \citet[][light blue]{aab+21} and the minimum scattering times of the complete sample of 40 CHIME/FRB-discovered repeating sources of FRBs (light orange) on a log scale. Histograms of scattering times after discarding censored values are overplotted (dark blue for nonrepeaters and dark orange for repeaters).}
\label{fig:min_scat_repvsoneoff}
\end{figure}

\paragraph{DM Comparisons}

Figure \ref{fig:dm_oneoffvsrepeater} compares the DM distribution of all one-off events in \citet{aab+21} and the complete sample of CHIME/FRB-discovered repeating sources of FRBs. Our results suggest a significant difference in the DM distributions of repeaters and nonrepeaters. The AD and KS tests give $p_{\textrm{AD}} = 0.0112$ and $p_{\textrm{KS}}=0.00618$, which suggest observed nonrepeaters and repeaters come from different distributions. The analogous analysis previously reported by \citet{aab+21} using a smaller repeater sample detected no such difference. To investigate the putative difference further, we examined how the $p$-values change as we lower the S/N constraint to 11 and 10, in order to increase the sample sizes, though at the possible expense of completeness, but noting that the latter applies both to repeaters and nonrepeaters. When the S/N threshold is set to $>$ 11, $p_{\textrm{AD}} = 0.00425$ and $p_{\textrm{KS}}=0.00324$. For S/N $>$ 10, $p_{\textrm{AD}} =0.00196$ and $p_{\textrm{KS}}=0.00278$. Thus, we see that the differences become more significant at lower S/N thresholds where the sample sizes increase, as expected for a significant difference. For lower S/N cuts, the difference in mean DMs between repeaters and nonrepeaters also increases. At S/N $>$ 12, repeaters have a mean DM of 500 $\pm$ 50\,pc\,cm$^{-3}$ and nonrepeaters have a mean DM of 662 $\pm$ 24\,pc\,cm$^{-3}$, with a difference of 162 $\pm$ 55 pc cm$^{-3}$ between repeaters and nonrepeaters. At S/N $>$ 11, repeaters have a mean DM of 493 $\pm$ 47\,pc\,cm$^{-3}$ and nonrepeaters have a mean DM of 668 $\pm$ 22\,pc\,cm$^{-3}$, resulting in a difference of 175 $\pm$ 52\,pc\,cm$^{-3}$ (means differ with $>99\%$ confidence). At S/N $>$ 10, repeaters have a mean DM of 495 $\pm$ 45\,pc\,cm$^{-3}$ and nonrepeaters have a mean DM of 692 $\pm$ 23\,pc\,cm$^{-3}$, giving a difference of 197 $\pm$ 50\,pc\,cm$^{-3}$ (means differ with $>99\%$ confidence). 

This result is also observed in eDM distributions, as shown in Figure \ref{fig:edm_oneoffvsrepeater}. With our nominal cuts on the sample, we see a possible difference in the underlying distributions, with $p_{\textrm{AD}} = 0.0105$ and $p_{\textrm{KS}}=0.00416$.  The difference becomes more significant with lower S/N constraints (for S/N $>$ 11, $p_{\textrm{AD}} = 0.00464$ and $p_{\textrm{KS}} = 0.00465$; for S/N $>$ 10, $p_{\textrm{AD}} = 0.00188$ and $p_{\textrm{KS}} = 0.00204$), again supporting the existence of a difference in DM distributions between repeaters and apparent nonrepeaters. As the S/N threshold is lowered, the difference in mean eDM increases as well. For S/N $>$ 12, repeaters have a mean eDM of 436 $\pm$ 49\,pc\,cm$^{-3}$ and nonrepeaters have a mean eDM of 597 $\pm$ 24\,pc\,cm$^{-3}$, with a difference of 161 $\pm$ 55\,pc\,cm$^{-3}$ between repeaters and nonrepeaters. At S/N $>$ 11, repeaters have a mean eDM of 431 $\pm$ 46\,pc\,cm$^{-3}$ and nonrepeaters have a mean eDM of 602 $\pm$ 22\,pc\,cm$^{-3}$, with a difference of 171 $\pm$ 51\,pc\,cm$^{-3}$ (means differ with $> 99\%$ confidence). At S/N $>$ 10, repeaters have a mean eDM of 431 $\pm$ 44\,pc\,cm$^{-3}$ and nonrepeaters have a mean eDM of 627 $\pm$ 23\,pc\,cm$^{-3}$, giving a difference of 196 $\pm$ 50\,pc\,cm$^{-3}$ (means differ with $> 99\%$ confidence).

We thus conclude that with the larger sample of repeaters reported here, we detect a significant difference in DM distribution between repeaters and apparent nonrepeaters, with repeater DMs systematically lower than those of apparent nonrepeaters. This difference persists when we also include the ``silver'' sample with the ``gold'' sample (events with $R_\mathrm{cc} <5 $) while keeping the same nominal cuts restrictions.
The possible reasons for this difference are discussed below in \S\ref{sec:discussion}.

\paragraph{Scattering Time Comparisons} 
\label{sec:scattering_comparisons}

Next, we look at the distributions of scattering times.  As many of our scattering measurements are upper limits, we make use of statistical survival tests to consider differences in distributions. The Peto \& Peto and log-rank tests for the repeater versus apparent nonrepeater data indicate similar underlying distributions, with $p = 0.2$ and $p = 0.09$, respectively. Since large scattering times may be due to confusion with multiple sub-bursts merging together, making it hard to identify each sub-burst individually, we also compare the lowest scattering time of each repeating source with the scattering time of apparent nonrepeaters (see Fig.~\ref{fig:min_scat_repvsoneoff}). We find that the Peto \& Peto test yields $p = 0.1$, and the log-rank test yields $p = 0.07$, respectively, which shows there is no significant difference in scattering time distributions in the present samples. Even as we increase the sample size by lowering the S/N threshold to 10, we observe no significant difference (i.e., $p > 0.01$ for the log-rank and the Peto \& Peto test). Using NADA's \texttt{cenfit} function\footnote{\url{https://rdrr.io/cran/NADA/man/cenfit.html}} \citep{helsel2005nondetects,HarringtonFleming1982,Rsoftware,NADApackage}, we find the Kaplan-Meier estimate mean, which considers upper limits, of the lowest scattering time of repeaters to be 2.32 $\pm$ 0.63\,ms and that of nonrepeaters to be 3.22 $\pm$ 0.50\,ms.

\begin{table}[t]
\begin{center}
\caption{Summary of Repeater vs Apparent nonrepeater Distribution Comparisons for Nominal  Cutoff Criteria}
\begin{tabular}{lcccc} \hline
    Property & $p_{AD}^{a}$  & $p_{KS}^{b}$  \\\hline
    Right Ascension & 0.322 & 0.492 \\
    Declination   & 0.827 & 0.914  \\
    DM  & 0.0112 &  0.00618 \\
    eDM$^{c}$& 0.0105 & 0.00416  \\
    {\tt bonsai} S/N  &  0.262 & 0.405  \\
    {\tt fitburst} S/N & 0.691 & 0.855  \\
    Boxcar width   & 2.49$\times$ 10$^{-5}$ & 6.20 $\times$ 10$^{-5}$ \\
    Bandwidth  & 1.24 $\times$ 10$^{-8}$ &5.35$\times$ 10$^{-8}$ \\
    \hline
    Property  & Peto \& Peto  & Log-Rank   \\\hline
    Fluence$^d$  & 0.409 & 0.217 \\
    Flux$^d$ & 0.339 & 0.997\\
     Width$^e$ &  1.88 $\times$ $10^{-8}$ & 3.73 $\times$ $10^{-5}$  \\
    Scattering$^e$ & 0.222 & 0.0859\\ 
    Lowest Scattering$^{e,f}$ & 0.140 & 0.0674 \\
    
    \hline
\end{tabular}
\label{ta:rep_vs_nonrep}
\end{center}
    \textbf{Notes.} \\
    $^a$Anderson-Darling probability of originating from same underlying population.\\
    $^b$Kolmogorov-Smirnov probability of originating from same underlying population.\\
    $^c$Extragalactic DM. \\
    $^d$Incorporates lower limits --- see \S\ref{sec:comparison}. \\
    $^e$Incorporates upper limits --- see \S\ref{sec:comparison}. \\
    $^f$Uses only lowest scattering time per repeater. 
\end{table}

\subsection{Burst Rate Analysis}
\label{sec:rate_analysis}

Here we compute burst repetition rates for all our repeaters, putting them into the broader context of FRB burst rates.

To compute repetition rates, we first determine the on-sky exposure for each of the sources from the start of the experiment, 2018 August 28, through to 2021 May 1. We follow the exposure determination techniques previously used by \citet{aab+21}. In summary, we generate an all-sky exposure map for the aforementioned interval by querying metrics which characterize the up-time and sensitivity of the CHIME/FRB system. These metrics are combined with the CHIME/FRB beam model, under the assumption that a sky location is detectable if it is within the FWHM region of a synthesized beam at 600\,MHz. We query the resulting exposure map over a grid of positions within the 90\% confidence localization region for each source. The mean and standard deviation of the exposure over the corresponding positional grids are provided in Table \ref{tab:sources} and Table \ref{tab:candidate_sources}, for the ``gold'' and ``silver'' samples, respectively. The exposure increases with declination, with the uncertainties in the exposure dominated by the corresponding source declination uncertainties. 

\begin{figure}
\centering
\includegraphics[width=\textwidth]{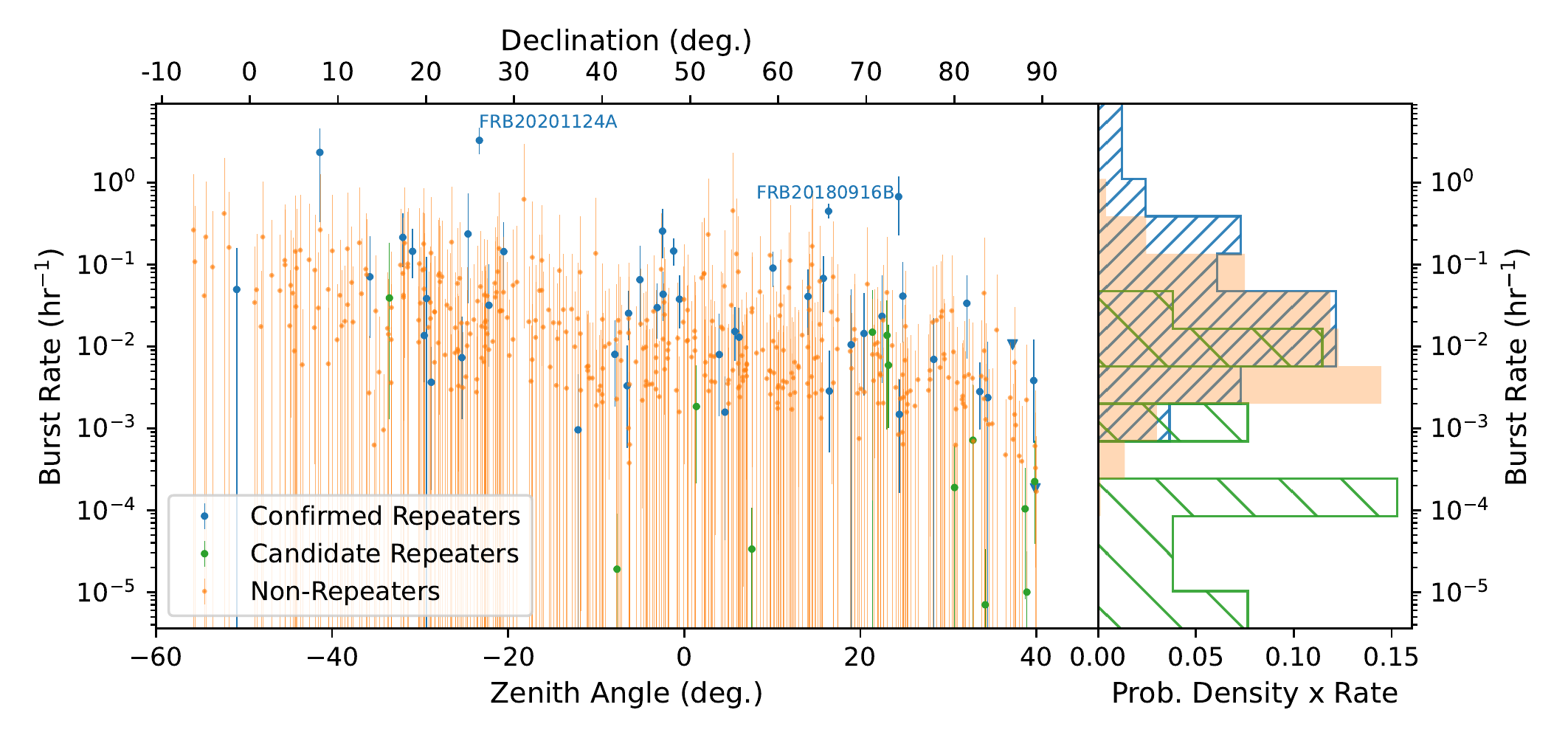}
\caption{Burst rates of previously published and new repeaters, candidate repeaters (the ``silver" sample) and Catalog 1 nonrepeaters as a function of zenith angle (bottom abscissa axis) and declination (top abscissa axis). For each source, the uncertainty on the rate is the quadrature sum of the 90\% Poisson uncertainty and the uncertainty in the source exposure (see Tables \ref{tab:sources} and \ref{tab:candidate_sources}). Histograms of the rate estimates for each class of sources are plotted in the right panel.}
\label{fig:repeaterburstrate}
\end{figure}

In the interval used for the exposure calculation, we detected a total of 128 bursts from the 25 confirmed and 14 candidate repeaters presented in this paper (i.e., bursts that were detected outside the FWHM at 600\,MHz of a synthesized beam were excluded from the burst rate computation). While calculating the burst rate, we exclude five of these bursts which were detected when the system was not operating nominally. Additionally, for circumpolar sources ($\delta > +70^\circ$), we exclude bursts detected in the lower transit. Therefore, we only compute the burst rate in the upper transit for all sources. This choice is motivated by the sensitivity of the primary beam being greater in the upper transit, which can provide a stronger constraint on the source activity. The sensitivity to each source also varies based on the synthesised beam response. This variation is accounted for in the 90\% confidence completeness thresholds reported in Tables \ref{tab:sources} and \ref{tab:candidate_sources}. To facilitate comparison across repeaters, we scale all burst rates to a fluence threshold of 5\,Jy\,ms, i.e., the average sensitivity of the CHIME/FRB system \citep{jcc+21}. The scaling involves the assumption of a power-law index of $-1.5$ for the cumulative burst energy distribution of each source. We set the index to be $-1.5$ to be consistent with the repetition rate analyses presented by \citet{abb+19c} and \citet{fab+20}. The scaled repetition rates are plotted in Figure \ref{fig:repeaterburstrate}. 

We also update the rate estimates for 19 previously published CHIME/FRB repeaters \citep{abb+19b,abb+19c,fab+20,lac+22} by computing the exposure up to 2021 May 1 using the corresponding source localization regions (see Table~\ref{tab:newreprates} in Appendix~\ref{sec:newreprates}). For 13 of these sources, we use unpublished baseband localization regions (Michilli et al., in prep.). The localization regions published in the discovery papers are used for the rest of the sample. A total of 210 bursts (including unpublished detections) were observed in the upper transits of these sources, of which we exclude 15 bursts as they were detected in low-sensitivity periods. We scale the repetition rates to a threshold of 5\,Jy\,ms, using the completeness thresholds reported in the discovery papers. To allow for comparison of the CHIME/FRB repeaters with the nonrepeaters, we also estimate burst rates for Catalog 1 one-off sources based on the detection of one burst from each source. Although \citet{aab+21} report the detection of 474 such sources, we exclude 63 of them because of their detection in low-sensitivity periods or in the lower transit. Another 14 sources are excluded as they are found to be repeat bursts from the sources presented in this work. The rates for the remaining sources are estimated using the header localization regions and completeness thresholds reported by \citet{aab+21}, and are plotted in Figure \ref{fig:repeaterburstrate}. 

Next we investigate whether the burst rates for the repeating sources are correlated with other properties, namely, extragalactic DM, intrinsic width, bandwidth, fluence and scattering. We perform this analysis with the confirmed and candidate repeaters presented in this work as well as the 19 previously published repeaters. For these 19 repeating sources, we use the burst properties reported in their discovery papers. Additionally, we determine properties of the unpublished detections from these sources using the techniques described in Section \ref{sec:props}. 

While testing for potential correlations, here we define eDM for a source as the mean extragalactic DM estimated using the NE2001 and YMW16 models. The source width is assumed to be the average of the intrinsic widths for all repeat bursts. If a repeat burst comprises of sub-bursts, we average the widths for the individual sub-bursts before using them to calculate the mean width for the source. We obtain the error on the mean width for each source as the quadrature sum of the uncertainties in the width measurements of the repeat bursts. We obtain mean bandwidths and fluences for each source and the corresponding errors in a similar manner. As for scattering, we adopt the strongest constraint among all repeat bursts as the scattering time for each source. This assumes that scattering structures along the line of sight to these sources do not change significantly in the interval between detections \citep[but see][]{occ+23}.
Due to the finite time resolution of the CHIME/FRB system, some width and scattering time measurements are upper limits. In such cases, we assume the 95\% confidence upper limit to be the measured value. 

\begin{figure}
\centering
\includegraphics[width=0.8\textwidth]{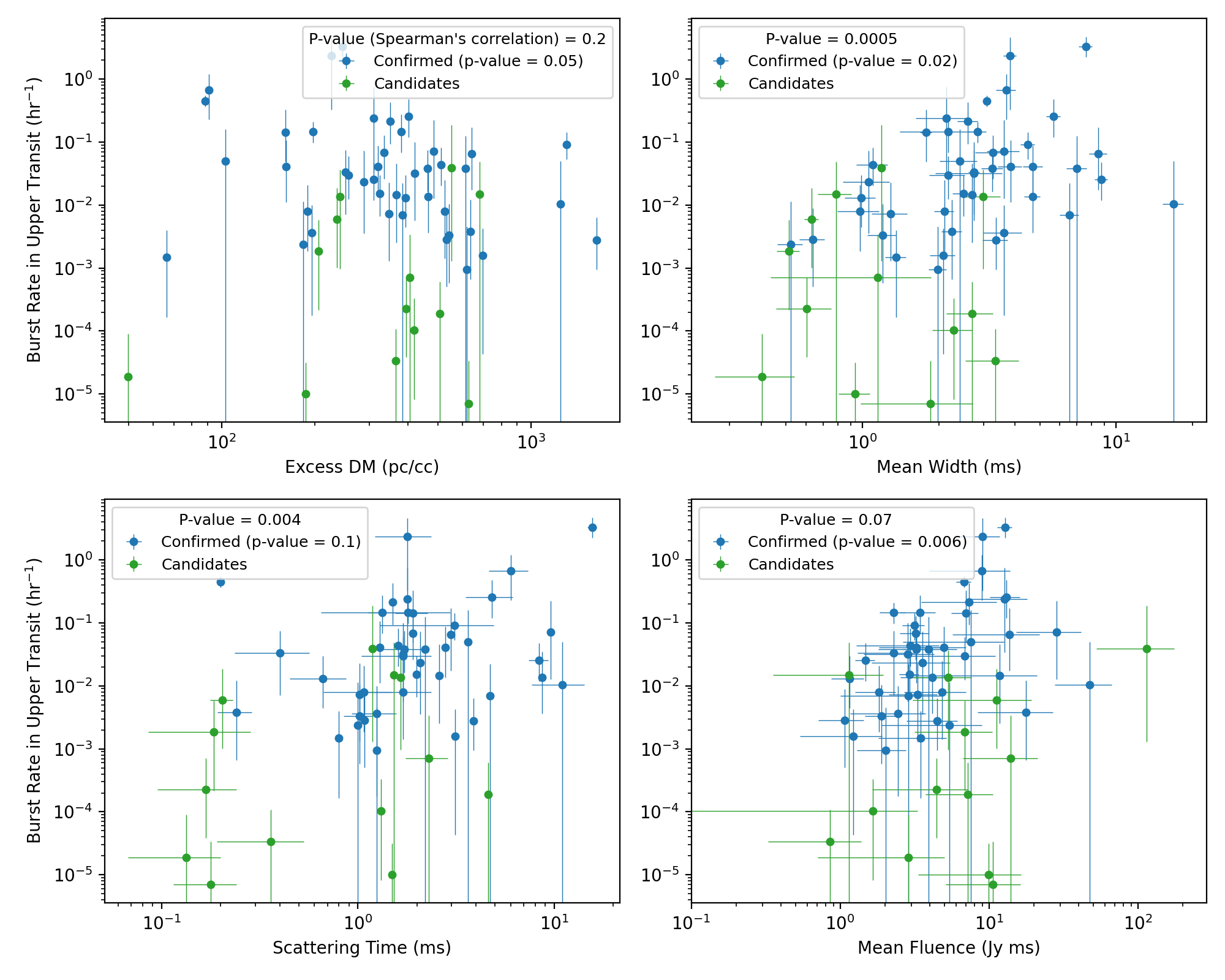}
\caption{Burst rates of previously published and new repeaters (blue) and candidate repeaters (the ``silver'' sample; green) as a function of eDM (top left), mean width (top right), scattering time (bottom left) and mean fluence (bottom right). The title of the legend in each panel indicates the associated $p$-value for the Spearman rank-order correlation (confirmed and candidate repeaters; blue and green), with the null-hypothesis being that the plotted quantities are uncorrelated. The $p$-value for the test conducted only using the sample of confirmed repeaters (blue) is indicated in brackets. The errors represent the 90\% confidence interval for the plotted quantities.}
\label{fig:ratecorrelations}
\end{figure}

While checking for a correlation of the repetition rate with these properties, we account for the errors in the measurements by performing 10,000 simulation runs. In each run, we sample a possible value of the burst rate and also of the property in consideration (say, the width) for each of the repeaters. For each repeater, we sample these values from a Gaussian distribution with a standard deviation equal to the 1$\sigma$ error on the burst rate and width. In each iteration, we check for a correlation using the Spearman's rank-correlation test. We then obtain the median of the $p$-value for the test over all 10,000 iterations. The burst rates and other properties along with the associated median $p$-values are shown in Figure \ref{fig:ratecorrelations}.

\section{Discussion}
\label{sec:discussion}

\subsection{Completeness of Repeater Identification}
\label{sec:completeness}

We believe that our repeater identification is complete for sources with properties similar to previously discovered repeaters (e.g., no extreme DM evolution). Next we discuss ways in which our $R_\mathrm{cc}$ calculation could be made more robust and sensitive in the future.

The clustering analysis (\S\ref{sec:clustering}) at the root of the repeater search presented here has used large tolerances in sky position and DM, $\sigma_\alpha = 2.2\degr / \cos(\delta)$, $\sigma_\delta = 0.5\degr$ and $\sigma_\mathrm{DM} = 13$\,pc\,cm$^{-3}$, that reflect uncertainties in the CHIME/FRB detection pipeline. It successfully constructs clusters of repeater detections in the main lobes and first sidelobes of the beams, but it will miss repeat bursts in the far sidelobes of the telescope, if they are detectable (this depends on the luminosity function of the repeaters, see also Lin et al. in prep.). The algorithm groups together events with $\Delta$DM $\lesssim 13$\,pc\,cm$^{-3}$, which makes it sensitive to mild putative DM evolution of repeating sources as well as to a range of DM measurement errors in the real-time detection pipeline. We allow for this range because in cases of downward-drifting subbursts, the DM can be overestimated by the S/N-optimizing detection pipeline as subbursts get superimposed for greater S/N \citep{hss+19}. Sources with DM evolution $\gtrsim 13$\,pc\,cm$^{-3}$ would have been missed, but this level of DM evolution has not been observed in any source of FRBs.

The $R_{\text{cc}}$ calculation is limited by a few observational realities. The probability that we detect an FRB event in some volume $\Delta\alpha\Delta\delta\Delta$DM is a function of CHIME/FRB's exposure and selection function (\S\ref{sec:pcc}). This is therefore not trivial to calculate. Instead, we empirically estimate it using a normalized histogram of our entire FRB sample, including only one burst from each known repeater. This means that the data set that is used to construct the null case for the $R_\mathrm{cc}$ calculation is contaminated by repeating sources. Ultimately, this produces conservative $R_{\text{cc}}$ estimates, but an improved method would use an uncontaminated null data set or account for the contamination in some way. We control for the problem of multiple hypotheses testing by multiplying by $N$, the total number of FRBs in this sample. This a common correction for the look-elsewhere effect, and it has been shown to be conservative \citep{gs14}.

Another factor that we do not account for is our pipeline-imposed increase in sensitivity for repeaters. As soon as an FRB candidate is detected we add it to a database of known sources. The S/N threshold for saving baseband data for any candidate is 12, but that threshold is lowered to 10 for events coincident with a known FRB source in the database. In principle, this should be accounted for in the empirical estimation of $p_{\{\alpha,\delta,\mathrm{DM}\}}$ but is non-trivial and will be the subject of future work.

The time between detections is also not accounted for in our $R_\mathrm{cc}$ calculation. Intuitively, FRBs with similar DMs and spatially coincident localizations that are detected in rapid succession (e.g., within a single CHIME/FRB transit) may seem more likely to be repeaters than those detected years apart. However, how to include this timing is not clear, particularly because we do not know the general temporal behavior of repeaters. More detailed quantitative characterizations of the waiting time distributions and other temporal properties of the repeater population could be helpful in this regard.

Future improvements to the $R_\mathrm{cc}$ framework could also allow for priors on additional parameters, such as the burst duration and bandwidths, which have been shown to be different for repeater bursts and one-off events \citep{pgk+21}, or the scattering properties, when we understand better what variability can exist \citep[see, e.g.,][]{occ+23}. Regardless, CHIME/FRB aims to provide sky positions with subarcminute precision for many future FRBs through its ``Outriggers'' upgrade \citep{lmm+21,clr+21}, which will make identifying true repeaters among the sample easier in the future.

\subsection{Comparison of New and Previously Known Repeater Samples}

As stated in \S\ref{sec:old_vs_new}, we conclude that there is no significant difference between new and previously published CHIME/FRB repeater samples. All $p$-values are $> 0.01$ except flux, which has a log-rank test $p$-value $< 0.01$ but has large uncertainty due to the presence of many censored values. We might expect to detect more repeaters in this new sample at lower declination where CHIME/FRB's sensitivity is lower, since more observation time has passed, but no difference in the declination distributions is observed. Similarly, due to longer total exposure, we might expect to see a difference in the DM distributions between previously published and new samples of repeaters since there is more opportunity to detect more distant repeaters, but no difference is observed. 

This lack of any apparent difference could indicate no difference in the two samples' properties, insufficient difference in exposure time for samples of these sizes to yield a significant detection, or too few sources to permit detection of any differences. Alternatively, the lack of an apparent difference between the new and previously published repeaters could be a result of FRBs having finite lifetimes, since the active lifetime of an FRB could end before a burst is detected. Or, new repeating sources (cf. FRBs~20201124A and 20220912) could ``turn on'' at all declinations and DMs at a rate comparable to the increase in discovery of sources at high DM and low declination with increased total exposure. The two competing rates could mask the expected increase in detections of more distant sources over time, especially when the total number of sources is still relatively low.

\subsection{Comparison of Apparent Nonrepeating and Repeating Sources}

As discussed in \S\ref{sec:rep_vs_nonrep} and shown in Table \ref{ta:rep_vs_nonrep}, the AD and KS tests indicate that the DM and extragalactic DM (eDM) distributions of the full repeaters sample and Catalog 1 apparent nonrepeaters (meeting the criteria specified in \S\ref{sec:rep_vs_nonrep}) are seemingly inconsistent with arising from the same population.  This inconsistency increases as we lower the S/N threshold. A likely explanation for the latter observation is that the sample size increases as we lower the S/N threshold, making any real differences more significant. If so, the difference in DM distribution between repeaters and apparent nonrepeaters is very significant. Such a difference could in principle indicate new, important support for the two source classes being of physically different origins.  A difference in DM distribution could indicate a difference in one or more of (i) distribution in redshift space, (ii) host galaxy type, or (iii) local environment. However, none of these is necessary to explain the effect. \citet{gcl+21} showed that the detection of nonrepeaters and repeaters is inherently subject to different observational selection effects. If repeaters and nonrepeaters originate from the same population, we would observe a difference in DM distributions because a second bright burst required for a source to be identified as a repeater is more challenging to detect at higher DM, if the luminosity function universally declines at higher luminosity.  Detailed population modelling will be required to determine whether the degree of the difference reported here can be explained by this simple bias assertion alone under reasonable assumptions of luminosity function, or whether one can conclude there is a genuine difference in source classes.

We do not observe a difference in scattering time distributions when we compare the first detected burst per source after the cuts stated in \S\ref{sec:rep_vs_nonrep}, nor do we observe a difference between the two distributions when we compare the lowest scattering time of each repeating source with nonrepeaters. We note that it is possible for  scattering-time measurements to be biased by broad pulse widths and complex morphology. Since repeater bursts have been shown to have longer duration and more complex morphology \citep{pgk+21}, we might expect a bias in which repeater bursts have longer scattering times compared to nonrepeaters. Yet, we observe repeaters to have comparable or even possibly lower scattering times both when we compare the first burst that passes the criteria  in \S\ref{sec:rep_vs_nonrep} (repeaters: 2.57 $\pm$ 0.77\,ms; nonrepeaters: 3.22 $\pm$ 0.50\,ms) and when we compare the minimum scattering time of each source (repeaters: 2.32 $\pm$ 0.63; nonrepeaters: 3.22 $\pm$ 0.50\,ms). On the other hand, if there were a strong DM/scattering correlation, then repeaters having a lower mean DM and eDM than apparent nonrepeaters, as we observe, would suggest they should also have lower scattering times. However, strong DM/scattering correlations have not been seen in the FRB population \citep[e.g.,][]{ckr+22}. Detailed population studies, preferably on burst properties measured using voltage data, for which the time resolution is higher, hence scattering may be more easily measured, will be needed to properly search our larger sample for such a correlation.

\subsection{Burst Rates}
\label{sec:reprates}

Here we discuss the burst rate analysis shown in \S\ref{sec:rate_analysis}, where we found burst rates for all new (both ``gold'' and ``silver'' samples) and previously published CHIME/FRB repeaters, as well as for Catalog 1 nonrepeating sources (see Fig.~\ref{fig:repeaterburstrate}).

First, we note that the estimated rates might not be equal to the true repetition rates of these sources because of two assumptions made in this analysis. First, assuming a power-law index of $-1.5$ for scaling the burst rates to a common fluence threshold will introduce rate errors of varying magnitudes for different sources. This is because the measured index for known repeaters varies widely, ranging from $-0.7$ to $-3.6$ \citep{lab+17,lac+22}. Second, some of the bursts in our sample could have been detected in the sidelobes of the synthesized beams. As the exposure map does not include exposure in the sidelobes, the true repetition rate for some sources could be lower than is reported here. However, as both these assumptions affect the estimated rates across all sources in a stochastic manner, they are unlikely to affect the conclusions of a rate comparison between the repeating and non-repeating populations and among the repeating sources. 

Among the repeating sources, we find that the candidate repeaters have, on average, lower repetition rates as only two bursts have been detected from most of these sources. We also find that the burst rate is, on average, lower for high-declination sources. This is because the increased exposure at high declinations allows us to detect repeating FRBs with low repetition rates and place stricter constraints on the rates for the one-off sources. While some active repeating sources, such as FRB~20201124A \citep{lac+22}, have anomalously high repetition rates, we find no clear bimodality in the rates between repeating and apparently non-repeating sources (right panel of Fig.~\ref{fig:repeaterburstrate}). Therefore, we cannot as yet rule out the possibility that all FRBs repeat (see also \S\ref{sec:doallfrbsrepeat} below). This conclusion is strengthened by the observation that some of the candidate repeaters have lower rates than the rate upper limits of one-off sources. The transit path for these candidate repeaters has higher sensitivity than the all-sky average, which results in the scaled detection rate (for a 5-Jy ms threshold) being lower than limits for one-off sources. The fact that we are preferentially seeing repeaters at sky locations with the highest sensitivity and exposure suggests that all FRBs could be seen to repeat, given enough exposure and sensitivity. It is beyond the scope of this paper to investigate how parameterizing clustered repetition would affect the results of this analysis, but would be an interesting future investigation.

In \S\ref{sec:rate_analysis}, we also searched for correlations between burst rate and various burst properties (see Fig.~\ref{fig:ratecorrelations}). The burst rate might be expected to be anti-correlated with eDM if the repeater luminosity function is, on average, steep, and if eDM is on average higher for more distance sources, both likely possibilities.  Higher rates for lower eDM under these assumptions would be expected because closer sources and their repeat bursts will be easier to detect. Although we see a slight hint of this for the confirmed repeaters ($p$-value = 0.05 for confirmed sources, but only 0.2 for the full sample; see Fig.~\ref{fig:ratecorrelations} top left), overall, the correlation is weak, suggesting more statistics may be needed to detect such an effect with high significance.

On the other hand, we see a significant correlation between average pulse width (defined as the average width of all detected bursts for each source) and burst rate ($p$-value = 0.00005 for the full sample, and 0.02 for just the confirmed sources; see Fig.~\ref{fig:ratecorrelations} top right). The CHIME/FRB instrumental selection function strongly disfavors detection of long bursts \citep{aab+21,mts+23}, so preferentially detecting more wide ones cannot be due to that bias.  On the other hand, CHIME/FRB's detection pipeline operates on data having time resolution of 0.983\,ms, so much narrower bursts, especially if faint, are selected against, a bias not made clear by the constant fluence injection model used when demonstrating our bias \citep[see][]{mts+23}. Hence, bursters having very narrow, frequent bursts \citep[e.g.,][]{nhs+23} are not well represented here.  Nevertheless, the correlation we detect is present in our data and may represent a genuine astrophysical trend. Such a correlation might be expected as longer-duration bursts may extend over wider solid angles, hence be more detectable \citep[e.g.,][]{con19}.

We also detect a possible positive correlation between burst rate and scattering time (here defined as the smallest scattering time among the repeats from each source) with $p$-value = 0.004 for the full sample, but only 0.1 for the confirmed sources (see Fig.~\ref{fig:ratecorrelations} bottom left). As rate is likely an intrinsic source property, whereas scattering time is likely related to the propagation path, a true correlation would be somewhat surprising if scattering originates anywhere but the environment local to the source.  On the other hand, such a correlation could arise if, for example, a more active source is preferentially embedded in a denser, more turbulent plasma. We note that highly scattered bursts are strongly selected against in our CHIME/FRB detection pipeline \citep{aab+21,mts+23}, so detecting more at large scattering time is unexpected given this instrumental bias.  On the other hand, many of our scattering times are upper limits (here treated as measurements), either because the time is smaller than our time resolution, or because it is smaller than the burst width, so the associated $p$-values must be treated as approximate at best.  Nevertheless, a possible correlation is present in our data but seems likely related to the rate/width correlation discussed above, since, again, detection of scattering is often strongly limited by pulse width.  We therefore do not ascribe great significance yet to this possible correlation and resist any astrophysical interpretation for the time being.

Finally, we detect a hint of positive correlation between burst rate and mean burst fluence ($p$-value 0.006 for confirmed sources and 0.07 for the full sample; see Fig.~\ref{fig:ratecorrelations} bottom right). This could be a selection bias; for example, closer sources will tend to have higher fluences, and hence be easier to detect repeatedly.  On the other hand, the absence of as strong a correlation between burst rate and eDM does not support that hypothesis. Alternatively, there could be a correlation between burst rate and luminosity, wherein the most active sources are also the most radiatively energetic. On the other hand, there could be no such correlation, yet we tend to detect the most luminous, active sources first.  If so, the top left corner of the plot will eventually fill out.

In summary, a preliminary study of correlations between repeater burst rate and various burst properties reveals some possible relationships that are intriguing. We note that in some cases, the correlations are significant with just the confirmed sample and sometimes with the full sample (i.e., including the ``silver'' sample presented here), suggesting that one or more of the putative correlations may be spurious.  CHIME/FRB will continue to detect repeating sources (see \S\ref{sec:frep} below), which will help improve the statistics. Detailed simulations accounting for the variety of instrumental and observational biases at play will be required before astrophysical conclusions can be drawn. However, these are underway.

\subsection{Repeater Fraction}
\label{sec:frep}

If all FRBs eventually repeat, the number of repeaters over the total number of observed FRB sources, $f_\mathrm{rep}$, will tend to an equilibrium close to unity as FRB surveys continue. The value of the equilibrium depends on the birth rate and lifetime of repeaters. If only some FRBs repeat, $f_\mathrm{rep}$ will plateau or turn over at a value significantly different from 1 \citep{agz21,gcl+21}. The evolution of $f_\mathrm{rep}$ furthermore depends on the luminosity function and repetition rate distribution of sources. To study $f_\mathrm{rep}$, we split our experiment into five declination bins with approximately equal exposure and sensitivity within each bin. We separate upper and lower transits around the NCP (bursts with circles and triangles as markers, respectively, in Fig.~\ref{fig:timeline}) which effectively results in six independent declination bins. This split is necessary because we want to investigate the dependence of the repeater detection rate on exposure and sensitivity; with a uniform distribution of repeaters on the sky, it is possible that we run out of new repeaters to detect more quickly at sky positions with more daily exposure.

To ensure completeness, we only include events in this analysis with detection S/N $>$ S/N$_\mathrm{threshold}$, with S/N$_\mathrm{threshold} \in \{10, 11, 12\}$. For this analysis, we ``discover'' a repeating source of FRBs the second time we detect from it a burst with S/N $>$ S/N$_\mathrm{threshold}$ (highlighted with an extra circle or triangle in Fig.~\ref{fig:timeline}). We count the number of FRBs, $N_\mathrm{FRB}$, and repeating source discoveries, $N_\mathrm{rep}$, up to each UTC day and assign Poisson counting errors ($\sigma_N \approx \sqrt{N}$, for large $N$, using \texttt{scipy.stats.poisson.interval} to be precise, also for small $N$). We extrapolate the FRB detection rate after the cutoff date for the first catalog (2019 June 30)  by approximating the detection rate as the total number of FRBs detected in each analysis bin divided by the exposure up until the cutoff date. 

The repeater fraction
\begin{align}
    f_\mathrm{rep} = \frac{N_\mathrm{rep}}{N_\mathrm{FRB}},
\end{align}
has uncertainty
\begin{align} 
    \sigma_{f_\mathrm{rep}} = \sqrt{ \left(\frac{\sigma_{N_\mathrm{rep}}}{N_\mathrm{FRB}}\right)^2 + \left( \frac{N_\mathrm{rep}\sigma_{N_\mathrm{FRB}}}{N_\mathrm{FRB}^2} \right)^2 }
\end{align}
from propagating errors. For the uncertainty calculations, we keep $N_\mathrm{FRB}$ and $\sigma_{N_\mathrm{FRB}}$ fixed to their measurements on 2019 June 30 instead of calculating uncertainties based on extrapolated values. We report the 90\% confidence interval.

\begin{figure} 
\centering
\includegraphics{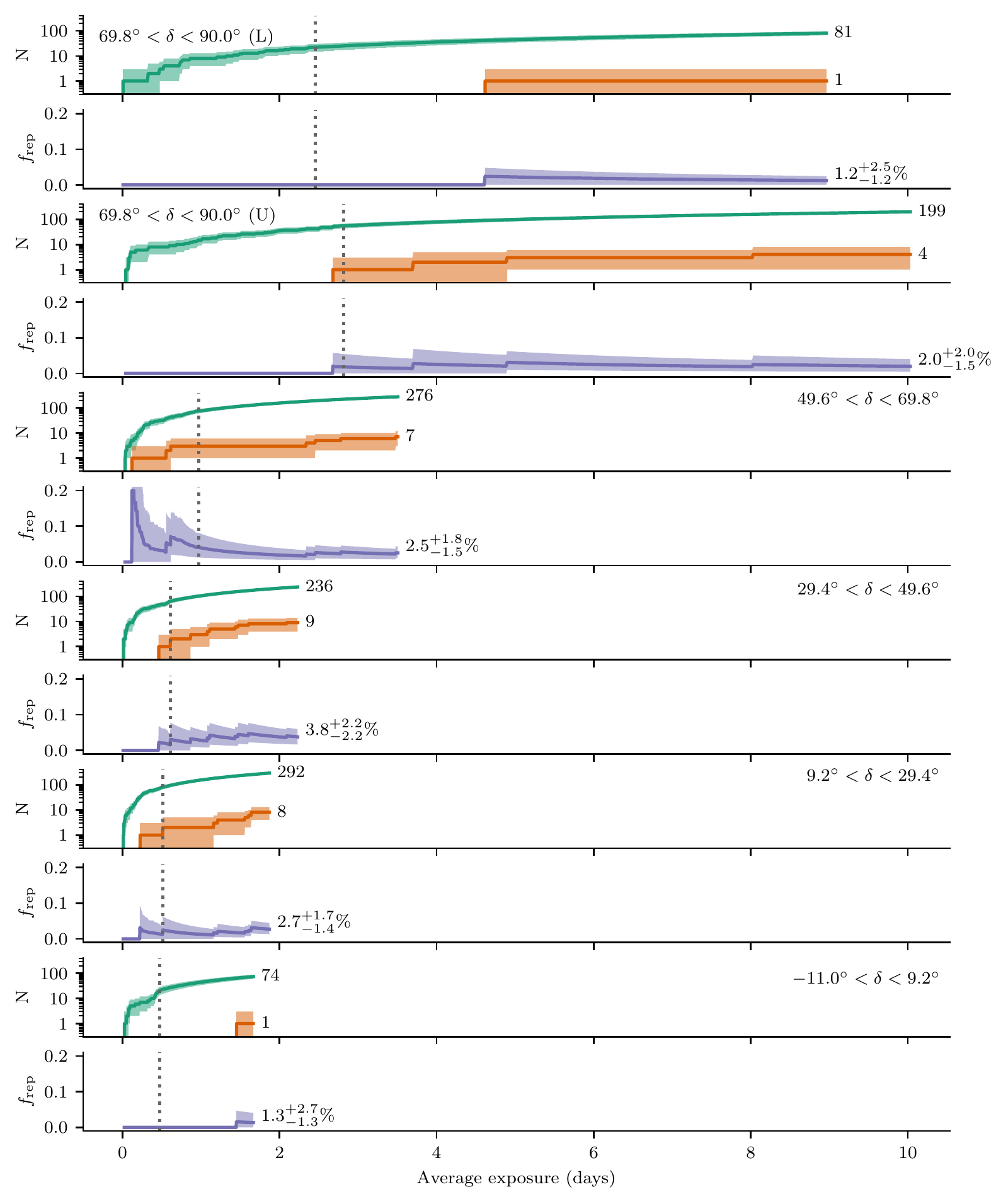}
\caption{Fraction of repeating FRBs over all FRB sources, $f_\mathrm{rep}$, as a function of average total exposure in different declination bins (ranges are indicated in the top left or right corners, with U/L separating the ``upper'' and ``lower'' transit of the top bin, respectively) with approximately equal exposure and sensitivity within each bin. In each pair of panels the first shows the number of  FRB (green) and repeater (orange) sources and the second value of $f_\mathrm{rep}$ (purple) with the values at 2021 May 1 written out (90\% confidence). The FRB rate is extrapolated after the cutoff date (gray dotted line) for the first catalog. This figure includes only detections with S/N $> 12$.}
\label{fig:frep}
\end{figure}

The repeater fractions are shown as a function of average exposure (in a given declination bin) in Figure~\ref{fig:frep} for a S/N $>$ 12 threshold and without imposing a cutoff of the maximum extragalactic DM. It can be seen that $f_\mathrm{rep}$ tends to an equilibrium of 1--4\% in all bins, with consistent values between bins at the 90\% confidence level. Summing FRBs and sources over all bins, we find an overall $f_\mathrm{rep} = 2.6_{-2.6}^{+2.9}$\%. The repeater fraction initially reached $\sim$20\% in the $49.6\degr < \delta < 69.8\degr$ bin because FRB 20180916B \citep{abb+19b} was discovered as a repeating source early on. The computing node that processes eight beams in the $27\degr < \delta < 33\degr$ range has been sporadically on and off during periods of the experiment. We have thus checked that excluding this declination range does not affect our results significantly, and have eventually included this declination range for the measurement presented here. We do not currently have the statistical power to differentiate between $f_\mathrm{rep}$ tending to an equilibrium or turning over. We have also inspected the FRB and repeater detection rate as a function of UTC time (not exposure, as in Fig.~\ref{fig:frep}) and find no evidence for a change of rate there either. The rates as a function of exposure and as a function of UTC time probe slightly different properties of the sources, as the total exposure in all cases is collected over a $\sim$three-year period.

\begin{figure}
    \centering
    \includegraphics[width=\textwidth]{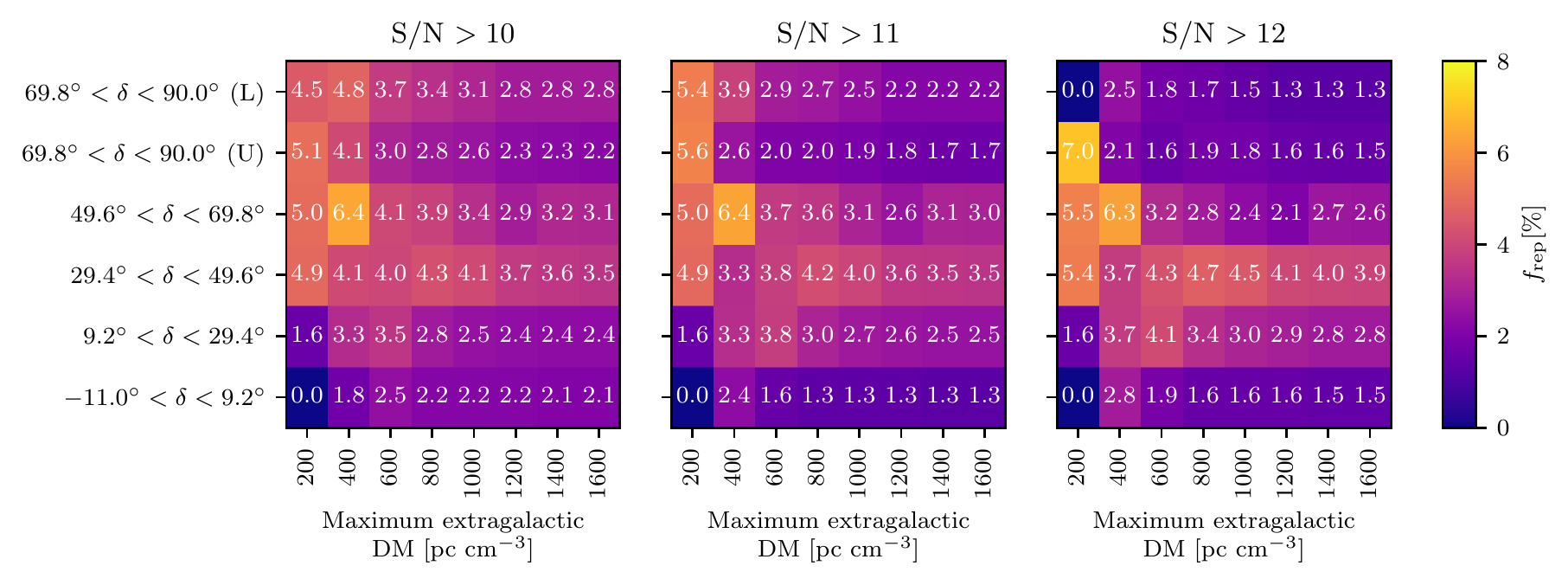}
    \caption{Evolution of $f_\mathrm{rep}$ (in \%) in six declination bins as a function of maximum extragalactic DM (NE2001) for three different S/N thresholds. While there are hints of trends towards higher $f_\mathrm{rep}$ at lower maximum extragalactic DM and closer to zenith ($\delta = $49.6\degr), note that uncertainties are not shown here and that all values are consistent at the 90\% confidence level.}
    \label{fig:frepevolution}
\end{figure}

We investigate the behavior of $f_\mathrm{rep}$ as we lower the S/N threshold and as we probe only a local volume of the Universe by imposing a maximum extragalactic DM for sources to include. The evolution is shown in Figure~\ref{fig:frepevolution}. At maximum extragalactic DM $<$ 200\,pc\,cm$^{-3}$, we have not yet discovered any repeating sources in some bins, likely due to small number statistics. In general, we observe that $f_\mathrm{rep}$ is higher at lower maximum extragalactic DM and gradually falls off as we probe higher extragalactic DMs (in line with findings in \S\ref{sec:rep_vs_nonrep} that repeaters have lower mean DM and extragalactic DM). However, in all cases, the repeater fractions in different declination bins are consistent at the 90\% confidence level. We observe a higher $f_\mathrm{rep}$ in the two bins around CHIME's zenith ($\delta = $49.6\degr) where our sensitivity is highest. This could just be a selection effect from being more sensitive to fainter repetitions and thus detecting repeaters at a higher rate. The trends are similar regardless of S/N threshold and the use of the NE2001 or the YMW16 model for the estimation of the extragalactic DM.

\citet{smb+23} used a detailed population synthesis to infer a volumetric rate of of $7.3 \times 10^4$ bursts\,Gpc$^{-3}$\,year$^{-1}$ above a pivot energy of $10^{39}$\,erg and below a scattering timescale of 10\,ms at 600\,MHz based on FRBs in CHIME/FRB's first catalog. Based on the observed $f_\mathrm{rep}$ in CHIME/FRB, the rate of active repeating sources can only be a few percent of that rate. Without an analogous population synthesis accounting for the observed $f_\mathrm{rep}$, we cannot make any firm conclusions about the statistics of the population of repeating FRBs versus a putative distinct population of truly one-off events. It is also not possible to compare $f_\mathrm{rep}$ as measured for the CHIME/FRB survey with $f_\mathrm{rep}$ in other surveys because other repeating sources of FRBs were discovered in targeted follow-up observations rather than untargeted observations \citep{ssh+16a,kso+19,lwm+20,ksf+21,nal+22}.

\newpage
\subsection{Do All FRBs Repeat?}
\label{sec:doallfrbsrepeat}

Even though multiple lines of evidence point to observed differences between the properties of repeater bursts and apparently one-off events, it is not yet possible to rule out that all FRBs repeat. Detailed population synthesis studies are necessary to properly interpret observations and rule out selection effects as the origins of the observed differences.

There is an observed dichotomy between the burst durations and bandwidths of repeater bursts and apparently one-off events \citep{pgk+21} that persists in this new sample (\S\ref{sec:comparison}). To this, we add an observed difference in DM distributions (\S\ref{sec:rep_vs_nonrep}) but no detectable bimodality in repeat rates (\S\ref{sec:reprates}). We find a significant correlation between burst width and repetition rate only if we include candidate repaters in the analysis, which might point to an evolution of burst properties with rate (\S\ref{sec:reprates}). Only 2.6\% of sources detected by CHIME/FRB has so far been seen to repeat.

Reconciling the dichotomy in burst morphology with one population of repeating FRBs with a continuum of repetition rates needs a correlation between repetition rate and burst duration (potentially observed; \S\ref{sec:reprates}) \emph{and} anti-correlation between repetition rate and bandwidth (not yet observed). This could either be achieved intrinsically through the emission mechanism or extrinsically through a propagation effect (e.g., tied to an evolutionary stage if active repeaters are young sources). Beaming geometry could provide a correlation between repetition and duration as wider opening angles could make bursts from sources both easier to detect and longer in duration \citep{cmg20}. However, it is unclear how a wider opening angle could also lead to narrower emission bandwidths.

Studies of the host galaxies \citep{bha+22} and galaxy offsets \citep{mfs+21} of FRBs have not found clear differences between those of repeating sources and one-off events, except that the hosts of repeaters FRBs~20121102A and 20190520B are outliers in the currently known population. 

An additional future probe of differences could be polarization properties, as more polarization properties of one-off events become available.  On the other hand, there already exists an observed wide diversity in the polarization properties of the repeaters \citep[e.g.,][]{mgm+22b}.

Matching the observables to an astrophysical population of FRBs requires population synthesis. This has been used previously to constrain the repetition statistics and energy function of repeaters \citep{csr+19,jof+20a,jof+20b,lpw20L} and, for CHIME/FRB data, to interpret the scattering timescales of FRBs in the first catalog \citep{ckr+22}. A publicly available general framework for population synthesis is provided by \texttt{frbpoppy}\footnote{\url{https://github.com/davidgardenier/frbpoppy}} \citep{gcl+21}. Here, we outline some considerations for accurately modeling FRBs with CHIME/FRB observations for future population studies. 

When modeling repeaters and one-off events as distinct populations, many new variables are introduced. Apart from separate volumetric rates, it needs to be carefully considered whether they share other population model parameters, such as those describing their energy function, evolution through cosmic time, and spectral properties. \citet{smb+23} fit an energy function and volumetric rate to CHIME/FRB's first catalog. As $\sim$90\% of FRBs in the first catalog are one-off events, it is likely that those measurements could be used well to model one-off events. Those and other model fits are, however, sensitive to the inclusion of repeating sources for which only one event has been detected as apparent nonrepeaters. There is also the question of how to model the burst rates of repeaters, as varying levels of burst activity have been observed across the population of identified repeating FRBs, including highly clustered and rarely repeating sources \citep[\S\ref{sec:reprates}; see also, e.g.,][]{oyp18,gcf+23,nhs+23}. Population synthesis will need to reproduce the observed distribution of repetition rates, while also reproducing the observed differences in other burst properties. 

In addition to modeling the underlying intrinsic population, consideration also needs to be taken towards modeling the survey in which these bursts were detected and comparing the modeled population bursts with the observed data. For this, modeling selection effects is crucial. To accurately forward-model the CHIME/FRB survey, it is important to take into account its exposure\footnote{The exposure up to 2021 May 1 is available here: \url{https://www.canfar.net/storage/vault/list/AstroDataCitationDOI/CISTI.CANFAR/23.0004/data/exposure}} and to use a beam model.\footnote{An early version of our beam model, used for injections, is publicly available here: \url{https://github.com/chime-frb-open-data/chime-frb-beam-model}} The CHIME/FRB synthesized beams have chromatic sensitivities  set by the primary beam response of the CHIME telescope and the beamforming algorithm, which can lead to, e.g., apparent, observed, bandwidths that are smaller than intrinsic bandwidths if bursts are detected away from the centers of beams \citep[see, e.g., \S2.1.3 and Fig.~2 in][]{pgk+21}. The selection function of the CHIME/FRB detection pipeline has been characterized with injections and has been made publicly available \citep{mts+23}.\footnote{\url{https://www.canfar.net/citation/landing?doi=22.0005}} The strongest bias of the detection pipeline is against bursts with scattering timescales above 10\,ms at 600\,MHz \citep{aab+21}. 

\section{Conclusions}
\label{sec:conclusions}

We have presented \nsources~new repeating sources of FRBs, plus \nappsources\ additional candidate repeaters, from a complete search of the CHIME/FRB database from the start of operations on 2019 September 3 up to 2021 May 1.  This brings the total number of repeating FRBs discovered by the CHIME/FRB project to 46, or counting the candidates, 60.  This large bounty is a result of CHIME's large field-of-view, high sensitivity, and daily surveying of a large area of sky.  Additional repeater discoveries continue to be made by CHIME/FRB at a steady pace.  

The DMs of the confirmed sources reported on in this paper fall in the range $\sim$200--1700\,pc\,cm$^{-3}$. Combining these with the DMs of previously reported CHIME/FRB repeaters, we have now demonstrated that on average, the DMs of repeaters are lower than those of the apparently nonrepeating sources found by CHIME/FRB.  This seems likely to be at least partly an observational bias, since, for reasonable luminosity functions, nearby repeaters will be found preferentially.  However, it may also signal a distinct astrophysical origin for repeaters versus nonrepeating sources; detailed population synthesis studies that account for the many observational and instrumental biases will be required to disentangle such effects. The possibility of distinct physical origins is supported but unproven by statistically different dynamic spectra between the two groups -- wider bursts and narrower spectra for repeaters --  as previously shown \citep{aab+21,pgk+21}, a result which is strongly supported by our analysis of our now much larger sample.

Interestingly, we have detected no significant bimodality in the repeat rates of repeaters and apparent nonrepeaters, although the most active CHIME/FRB repeaters do have anomalously high rates compared to the bulk of the studied population. While we measure that $2.6_{-2.6}^{+2.9}$\% of sources detected by CHIME/FRB repeat, the lack of bimodality in burst rates between the populations suggests that some fraction of our apparent nonrepeaters may yet repeat and that we cannot rule out all FRBs eventually repeating.

We encourage radio and multi-wavelength follow-up of the newly presented sources, and especially of our candidates.  As for previously published repeaters, we update a webpage\footnote{\url{https://www.chime-frb.ca/repeaters}} with new detections of these sources to communicate which sources are active while also continuing to provide the community with realtime events via VOEvents.\footnote{\url{https://www.chime-frb.ca/voevents}}

\begin{acknowledgements}

We acknowledge that CHIME is located on the traditional, ancestral, and unceded territory of the Syilx/Okanagan people.

We thank the Dominion Radio Astrophysical Observatory, operated by the National Research Council Canada, for gracious hospitality and expertise.

CHIME is funded by a grant from the Canada Foundation for Innovation (CFI) 2012 Leading Edge Fund (Project 31170) and by contributions from the provinces of British Columbia, Qu\'{e}bec and Ontario. The CHIME/FRB Project is funded by a grant from the CFI 2015 Innovation Fund (Project 33213) and by contributions from the provinces of British Columbia and Qu\'{e}bec, and by the Dunlap Institute for Astronomy and Astrophysics at the University of Toronto. Additional support was provided by the Canadian Institute for Advanced Research (CIFAR), McGill University and the Trottier Space Institute via the Trottier Family Foundation, and the University of British Columbia. The Dunlap Institute is funded through an endowment established by the David Dunlap family and the University of Toronto. Research at Perimeter Institute is supported by the Government of Canada through Industry Canada and by the Province of Ontario through the Ministry of Research \& Innovation. The National Radio Astronomy Observatory is a facility of the National Science Foundation (NSF) operated under cooperative agreement by Associated Universities, Inc.

\allacks

\end{acknowledgements}

\facilities{CHIME/FRB, TNS}

\software{Astropy \citep{astropy1,astropy2,astropy3}, bitshuffle \citep{mas17}, cythong \citep{bbc+11}, hdf5 \citep{hdf5}, HEALPix \citep{ghb+05}, healpy \citep{healpy}, Matplotlib \citep{matplotlib}, NADA \citep{NADApackage}, NumPy \citep{numpy}, pandas \citep{mck10,pandas}, SciPy \citep{scipy}, Uncertainties (\url{https://pythonhosted.org/uncertainties/})}

\bibliographystyle{aasjournal}

\bibliography{frbrefs}

\appendix

\section{Candidate repeating sources of fast radio bursts}
\label{sec:candidates}

\begin{figure}[ht]
    \centering
    \includegraphics[width=0.9\textwidth]{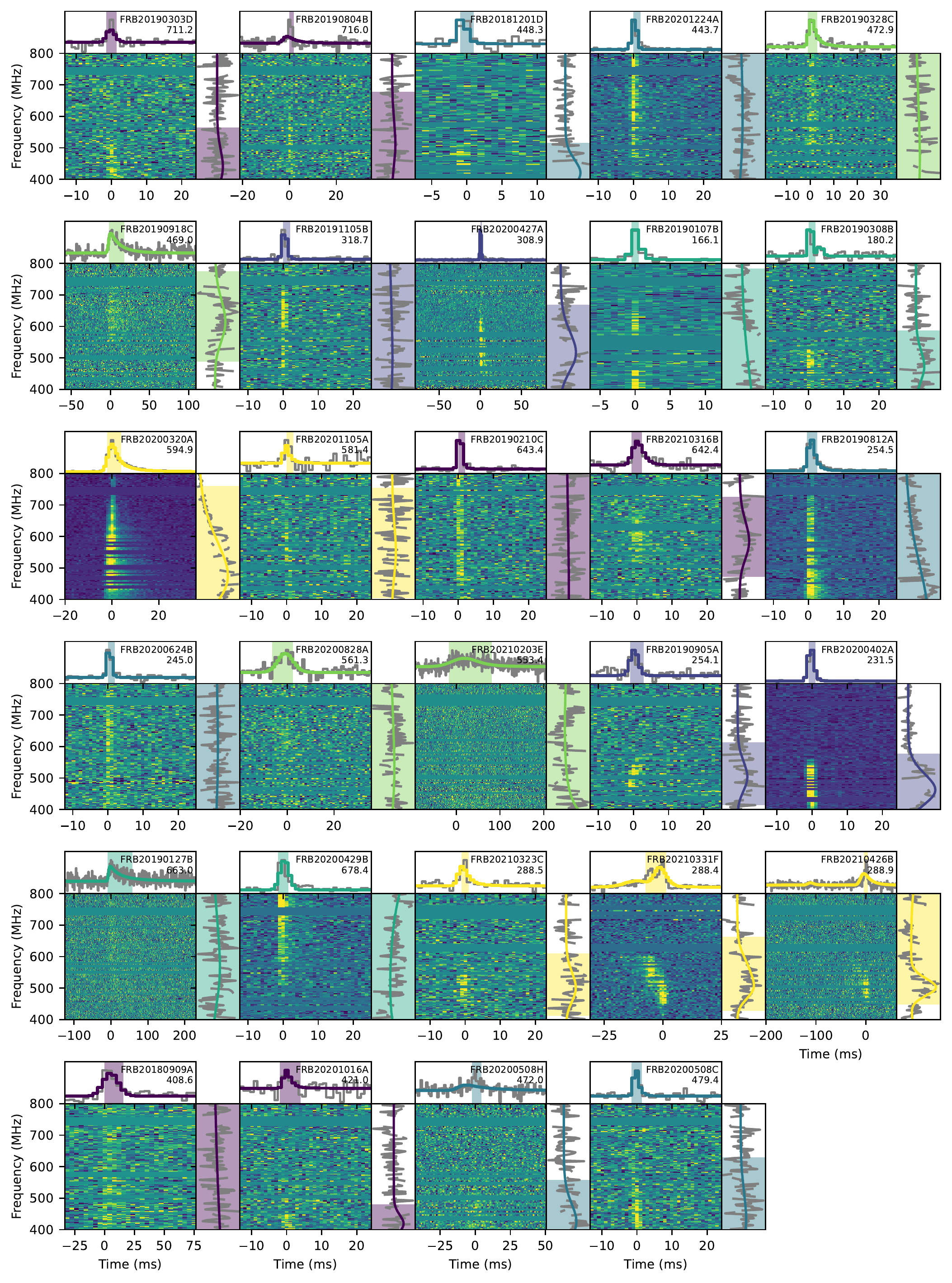}
    \caption{Same as Figure~\ref{fig:gold_wfalls_0}, but for candidate repeating sources of FRBs (i.e., ``silver'' sample).}
    \label{fig:silver_wfalls_0}
\end{figure}

Source properties of \nappsources~candidate repeating source of FRBs are summarized in Table~\ref{tab:candidate_sources}. The dynamic spectra of their bursts are shown in Figure~\ref{fig:silver_wfalls_0}. These candidates have lower significance, which by definition means the burst-to-burst DM and sky position differences are larger and the repetition rates are low. The DM differences are as large as $\sim$14\,pc\,cm$^{-3}$ (candidate FRBs~20190107B and 20200320A) and variation of this magnitude has not yet been observed in repeating sources of FRBs before. Some of the bursts in this candidate sample show larger than average bandwidths for repeater bursts, which we caution to overinterpret (i.e., we do not downweight their likelihood as candidates for this reason and we do not claim a correlation between bandwidth and repetition rate using this candidate sample). Some candidate sources also show apparently different scattering timescales from burst to burst (candidate FRBs~20190328C, 20201105A, 20200828A, 20210203E and 20190127B) that we cannot at this time confirm are real. Scattering variations have been detected in FRB~20190520B \citep{occ+23}.

\begin{table}[ht]
\begin{center}
\caption{Properties of \nappsources~New Candidate Repeating Sources of FRBs, Ordered by Increasing $R_\mathrm{cc}$ (Our ``Silver'' Sample)}
\centering
\resizebox{1.05\textwidth}{!}{ 
\hspace{-1.8in}
\begin{tabular}{ccccccccccccc} \hline
    FRB Source$^a$  &  $R_\mathrm{cc}$ & $\alpha^b$ & $\delta^b$ & $l^c$ & $b^c$ & DM$^d$ & DM$_{\rm NE2001}^e$ & DM$_{\rm YMW16}^e$ & N$_{\rm bursts}$  & Exposure$^f$ & Completeness$^g$ & Burst rate$^h$ \\
          &       &  (J2000)    &  (J2000) & (deg) & (deg) & (pc~cm$^{-3}$) & (pc~cm$^{-3}$) &(pc~cm$^{-3}$) &      &     (hr, upper / lower) &  (Jy ms) & (hr$^{-1}$) \\ \hline
20190303D & $0.62$ & 185.34(7) & 70.69(2) & $126.5$ & $46.2$ & 714.552(8) & 37 & 29 & 2 & $55 \pm 44$ / $142 \pm 3$ & 2.8 / 16.7 & $(1.49_{-1.71}^{+3.42}) \times 10^{-2}$ \\
20181201D & $0.64$ & 267(1) & 89.12(1) & $122.0$ & $27.4$ & 443.849(2) & 54 & 51 & 2 & $3591 \pm 288$ / $4359 \pm 14$ & 2.7 / 4.2 & $(2.24_{-1.85}^{+4.81}) \times 10^{-4}$ \\
20190328C & $0.71$ & $75.65_{-0.37}^{+0.10}$ & $82.11_{-0.19}^{+0.24}$ & $130.6$ & $23.3$ & 472.819(8) & 64 & 67 & 2 & $311 \pm 161$ / $294 \pm 161$ & 1.8 / 5.7 & $(7.16_{-7.74}^{+27.10})\times 10^{-4}$ \\
20191105B$^i$ & $0.73$ & 84.11(6) & 72.52(2) & $140.6$ & $20.4$ & 311.226(2) & 74 & 84 & 2 & $163 \pm 15$ / $161 \pm 0$ & 3.1 / 51.1 & $(5.9_{-4.9}^{+12.7}) \times 10^{-3}$ \\
20190107B$^j$ & $0.78$ & $49.31_{-0.11}^{+1.18}$ & $83.40_{-0.72}^{+0.13}$ & $127.1$ & $21.8$ & 168.253(3) & 68 & 73 & 2 & $362 \pm 189$ / $329 \pm 196$ & 0.2 / 0.4 & $(1.91_{-2.06}^{+7.20}) \times 10^{-5}$\\
 & & $22.37_{-0.80}^{+0.27}$ & $83.37_{-0.39}^{+0.08}$ & $124.1$ & $20.6$ & & 72 & 81 & & & \\
20200320A & $0.92$ & $42.45_{-0.04}^{+0.02}$ & $15.84_{-0.06}^{+0.07}$ & $160.2$ & $-38.3$ & 593.524(2) & 46 & 39 & 2 & $59 \pm 11$ & $8.7$ & $(3.91_{-3.78}^{+14.60}) \times 10^{-2}$ \\
20190210C & $1$ & $313.90_{-0.26}^{+0.13}$ & $89.19_{-0.47}^{+0.21}$ & $122.2$ & $26.7$ & 643.365(2) & 54 & 51 & 2 & $3354 \pm 1180$ / $3086 \pm 959$ & 1.7 / 2.6 & $<$ 1.83$\times 10^{-4}$ \\
20190812A & $1.4$ & $268.05_{-0.11}^{+0.22}$ & $50.71_{-0.02}^{+0.02}$ & $78.1$ & $29.8$ & 252.889(2) & 48 & 41 & 2 & $81 \pm 26$ & $0.9$ & $(1.85_{-1.63}^{+4.01}) \times 10^{-3}$\\
20200828A & $1.6$ & $114.08_{-0.20}^{+0.44}$ & $80.02_{-0.11}^{+0.09}$ & $134.1$ & $28.6$ & 555.41(2) & 54 & 50 & 2 & $241 \pm 133$ / $223 \pm 129$ & 0.4 / 1.6 & $(1.89_{-1.88}^{+4.20}) \times 10^{-4}$ \\
20190905A & $2.2$ & $73.27_{-0.03}^{+0.06}$ & $88.24_{-0.06}^{+0.07}$ & $124.6$ & $26.2$ & 231.5577(7) & 57 & 55 & 2 & $5399 \pm 980$ / $3928 \pm 1313$ & 0.4 / 0.6 & $(9.99_{-8.41}^{+21.50}) \times 10^{-6}$ \\
20190127B & $2.7$ & $169.26_{-0.08}^{+0.06}$ & $83.51_{-0.08}^{+0.08}$ & $126.0$ & $33.0$ & 678.350(3) & 48 & 42 & 2 & $345 \pm 191$ / $358 \pm 197$ & 0.1 / 0.3 &  (7$_{-8}^{+27}) \times 10^{-6}$ \\
20210323C & $3$ & $122.07_{-0.38}^{+0.11}$ & $72.35_{-0.14}^{+0.39}$ & $142.6$ & $31.6$ & 288.6(3) & 50 & 45 & 3 & $135 \pm 78$ / $108 \pm 60$ & 3.6 / 31.8 & $(1.37_{-1.27}^{+2.30}) \times 10^{-2}$ \\
20180909A & $3.9$ & $120.04_{-0.17}^{+0.30}$ & $57.00_{-0.07}^{+0.06}$ & $160.6$ & $31.7$ & 418.52(4) & 52 & 49 & 2 & $75 \pm 45$ & $0.1$ & $(3.35_{-3.41}^{+7.48}) \times 10^{-5}$\\
20200508H & $4$ & $54.60_{-0.27}^{+0.49}$ & $88.04_{-0.38}^{+0.32}$ & $124.4$ & $25.7$ & 479.430(6) & 58 & 57 & 2 & $1296 \pm 537$ / $1395 \pm 375$ & 0.8 / 1.1 & $(1.04_{-0.96}^{+2.27}) \times 10^{-4}$ \\
\hline
\end{tabular}
}
\label{tab:candidate_sources}
\end{center}
$^a$ Here we employ the TNS naming convention, see \url{https://www.wis-tns.org/astronotes/astronote/2020-70}. \\
$^b$ Positions were determined from baseband data where available and from per-burst S/N data otherwise (see \S\ref{sec:localization}), uncertainties are at 90\% confidence. \\
$^c$ Galactic longitude and latitude for the best position. \\
$^d$ Inverse-variance-weighted average DM, and the uncertainty in the calculation of that average. In cases where the DM was fixed in the \texttt{fitburst} fit, we set the DM uncertainty to 0.5\,pc\,cm$^{-3}$ for the calculation of the average. For candidates with large DM variation from burst to burst, this average and its uncertainty may not necessarily be representative and we advise looking at the individual burst DMs using generous DM ranges when conducting follow-up searches of these candidates. \\
$^e$ Maximum model prediction along this line-of-sight for the NE2001 \citep{ne2001} and YMW16 \citep{ymw17} Galactic electron density distribution models. Neither model accounts for DM contributions from the Galactic halo, which contributes no more than 111\,pc\,cm$^{-3}$ \citep{cbg+23}. \\
$^f$ For sources observed twice a day, the second entry corresponds to the less sensitive lower transit. The uncertainties in the total exposure for the upper and lower transits of each source are dominated by the corresponding source declination uncertainties since the widths of the synthesized beams vary significantly with declination. \\
$^g$ Fluence completeness limits are given at the 90\% confidence level. For sources observed twice a day, the second entry corresponds to the less sensitive lower transit. \\
$^h$ Adjusted to a 5\,Jy\,ms fluence threshold, assuming a $-1.5$ power-law energy index. \\
$^i$ The best-known sky position of this source falls in between the FWHMs at 600\,MHz of the synthesized beams. We report the exposure at the beam centre of the nearest beam and have scaled the fluence completeness threshold accordingly. \\
$^j$ Two equally likely ``islands'' in the localization 90\% confidence region.
\end{table}
\normalsize

\section{Burst Rates for Previously Published Repeaters}
\label{sec:newreprates}

\begin{table}[ht]
\begin{center}
\caption{Updated Burst Rates for Previously Published Repeaters}
\label{tab:newreprates}
\begin{tabular}{lccc} \hline
    FRB Source & Reference & Burst Rate$^a$  \\
     & & (hr$^{-1}$) \\ \hline
    20180814A & \citet{abb+19b} & $(6.57_{-4.47}^{+5.09})\times10^{-1}$\\
    20180916B & \cite{abb+19c} & $(4.48_{-0.86}^{+1.00})\times10^{-1}$ \\
    20181030A & \cite{abb+19c} & $(1.48_{-1.32}^{+2.47})\times10^{-3}$\\
    20181128A & \cite{abb+19c} & $(4.08_{-2.70}^{+4.63})\times 10^{-2}$ \\
    20181119A & \cite{abb+19c} & $(6.77_{-4.16}^{+5.95})\times10^{-2}$ \\
    20190116A & \cite{abb+19c} & $(3.18_{-2.62}^{+6.83})\times10^{-2}$ \\
    20181017A & \cite{abb+19c} & $(1.05_{-1.16}^{+3.97})\times10^{-2}$ \\
    20190209A & \cite{abb+19c} & $(6.95_{-6.96}^{+15.50})\times10^{-3}$\\
    20190222A & \cite{abb+19c} & $(1.44_{-1.19}^{+3.09})\times10^{-2}$ \\
    20190208A & \cite{fab+20} & $(4.34_{-2.30}^{+3.81})\times10^{-2}$\\
    20190604A & \cite{fab+20} & $(7.96_{-6.55}^{+17.10})\times10^{-3}$  \\
    20190212A & \cite{fab+20} & $(3.36_{-2.65}^{+4.08})\times10^{-2}$ \\
    20180908B & \cite{fab+20} & $(4.11_{-0.30}^{+6.52})\times10^{-2}$ \\
    20190117A & \cite{fab+20} & $(2.15_{-1.21}^{+2.09})\times10^{-1}$ \\
    20190303A & \cite{fab+20} & $(1.46_{-0.49}^{+0.64})\times10^{-1}$\\
    20190417A & \cite{fab+20} & $(9.05_{-3.71}^{+5.35})\times10^{-2}$\\
    20190213A & \cite{fab+20} & $(3.84_{-3.88}^{+8.56})\times10^{-2}$ \\
    20190907A & \cite{fab+20} & $(2.98_{-1.73}^{+2.93})\times10^{-2}$\\ 
    20200120E & \cite{bgk+21} & -- \\
    20201124A & \cite{lac+22} & 3.28$_{-1.06}^{+1.41}$ \\
    20171019A & \cite{kso+19} & -- \\ \hline
\end{tabular}
\end{center}
$^a$ Adjusted to a 5\,Jy\,ms fluence threshold, assuming a $-1.5$ power-law energy index. No rate measurements are reported for FRBs 20200120E and 20171019A as the sensitivity of the CHIME/FRB system to their source locations has not been measured (\S\ref{sec:reprates}).
\end{table}

\newpage

\section{Description of Fields in Catalog}
\label{sec:description}

In Table~\ref{ta:catalog descriptions}, we provide descriptions for each field present in our repeater catalog.

\begin{ThreePartTable} 
    \begin{longtable}[l]{c l l l}
    \caption{Repeater Catalog Field Descriptions} \endfirsthead
    \caption{\textit{continued}} \endhead
    \hline
    \\[-2ex]
    Column Number
    & Unit
    & Column Name
    & Description

    \\
    \\[-2ex]
    \hline
    \\[-2ex]
       0 & ... & \texttt{tns\_name} & TNS name \\
    1 & ... & \texttt{previous\_name} & Previous name (if applicable) \\
    2 &... & \texttt{repeater\_name} & Associated repeater name \\
    3 & degrees & \texttt{ra\_1} & Right ascension (J2000; first likelihood contour; see \S\ref{sec:localization}) \\
    4 & degrees & \texttt{ra\_1\_err\_low} & Right ascension lower error (see \S\ref{sec:localization}; first likelihood contour) \\
    5 & degrees & \texttt{ra\_1\_err\_up} & Right ascension upper error (see \S\ref{sec:localization}; first likelihood contour) \\
    6 & degrees & \texttt{ra\_2} & Right ascension (J2000; second likelihood contour; see \S\ref{sec:localization}) \\
    7 & degrees & \texttt{ra\_2\_err\_low} & Right ascension lower error (see \S\ref{sec:localization}; second likelihood contour) \\
    8 & degrees & \texttt{ra\_2\_err\_up} & Right ascension upper error (see \S\ref{sec:localization}; second likelihood contour) \\
    9 & ...& \texttt{ra\_notes} & Notes on right ascension \\
    10 & degrees & \texttt{dec\_1} & Declination (J2000; first likelihood contour; see \S\ref{sec:localization}) \\
    11 & degrees & \texttt{dec\_1\_err\_low} & Declination lower error (see \S\ref{sec:localization}; first likelihood contour) \\
    12 & degrees & \texttt{dec\_1\_err\_up} & Declination upper error (see \S\ref{sec:localization}; first likelihood contour) \\
    13 & degrees & \texttt{dec\_2} & Declination (J2000; second likelihood contour; see \S\ref{sec:localization}) \\
    14 & degrees & \texttt{dec\_2\_err\_low} & Declination lower error (see \S\ref{sec:localization}; second likelihood contour) \\
    15 & degrees & \texttt{dec\_2\_err\_up} & Declination upper error (see \S\ref{sec:localization}; second likelihood contour) \\
    16 & ... & \texttt{dec\_notes} & Notes on declination \\
    17 & degrees & \texttt{gl} & Galactic longitude  \\
    18 & degrees & \texttt{gb} &  Galactic latitude \\ 
    19 & hour& \texttt{exp\_up} & Exposure for upper transit of the source  \\
    20 & hour& \texttt{exp\_up\_err} & Exposure error for upper transit of the source  \\
    21 & ... & \texttt{exp\_up\_notes} & Notes on exposure for upper transit of the source  \\
    22 & hour& \texttt{exp\_low} & Exposure for lower transit of the source \\
    23 & hour& \texttt{exp\_low\_err} & Exposure error for lower transit of the source \\
    24 & ... & \texttt{exp\_low\_notes} & Notes on exposure for lower transit of the source  \\
    25 & ...& \texttt{bonsai\_snr} & Detection S/N  \\
    26 & pc cm$^{-3}$ & \texttt{bonsai\_dm} & Detection DM \\
    27 & Jy ms  & \texttt{low\_ft\_90} & Lower limit fluence threshold (90$\%$ confidence) \\
    28 & Jy ms  & \texttt{up\_ft\_90} & Upper limit fluence threshold (90$\%$ confidence) \\
    29 & Jy ms  & \texttt{low\_ft\_95} & Lower limit fluence threshold (95$\%$ confidence)\\
    30 & Jy ms  & \texttt{up\_ft\_95} & Upper limit fluence threshold (95$\%$ confidence)\\
    31 & ... & \texttt{snr\_fitb} & S/N determined using the fitting algorithm \fitburst \\
    32 & pc cm$^{-3}$ & \texttt{dm\_fitb} & DM determined using the fitting algorithm \fitburst\tnote{a} \\
    33 & pc cm$^{-3}$ & \texttt{dm\_fitb\_err} & DM error determined using the fitting algorithm \fitburst\tnote{a} \\
    34 & pc cm$^{-3}$ & \texttt{dm\_exc\_1\_ne2001} & \multirow{2}{*}{\parbox{3.8in}{DM excess between DM determined by \fitburst \ and NE2001 assuming the first likelihood sky position of the source}} \\
     & & &\\
    35 & pc cm$^{-3}$ & \texttt{dm\_exc\_2\_ne2001} & \multirow{2}{*}{\parbox{3.8in}{DM excess between DM determined by \fitburst \ and NE2001 assuming the second likelihood sky position of the source}} \\
     & & &\\
    36 & pc cm$^{-3}$ & \texttt{dm\_exc\_1\_ymw16} & \multirow{2}{*}{\parbox{3.8in}{DM excess between DM determined by \fitburst \ and YMW16 assuming the first likelihood sky position of the source}} \\
     & & &\\
    37 & pc cm$^{-3}$ & \texttt{dm\_exc\_2\_ymw16} & \multirow{2}{*}{\parbox{3.8in}{DM excess between DM determined by \fitburst \ and YMW16 assuming the second likelihood sky position of the source}} \\
     & & &\\
    38 & s & \texttt{bc\_width} & Boxcar width of the pulse (including all sub-bursts) \\
    39 & s & \texttt{scat\_time} & Scattering time at 600\,MHz\tnote{a} \\
    40 & s & \texttt{scat\_time\_err} & Scattering time error\tnote{a} \\
    41 & Jy & \texttt{flux} & Peak flux of the band-average profile (lower limit)  \\
    42 & Jy & \texttt{flux\_err} & Flux error \\
    43 & ... & \texttt{flux\_notes} & Notes on the burst flux \\
    44 & Jy ms & \texttt{fluence} & Fluence (lower limit) \\
    45 & Jy ms & \texttt{fluence\_err} & Fluence error \\
    46 & ... & \texttt{fluence\_notes} & Notes on the burst fluence \\
    47 & ... & \texttt{sub\_num} & \multirow{3}{*}{\parbox{3.8in}{Sub-burst number (if applicable). If the FRB has only one burst, then the sub-burst number is 0. Sub-bursts listed in chronological order. }}  \\
     & & &\\
     & & &\\
    48 & MJD & \texttt{mjd\_400} & \multirow{2}{*}{\parbox{3.8in}{Time of arrival with reference to 400.1953125 MHz for the specific sub-burst. }} \\
    & & &\\
    49 & MJD & \texttt{mjd\_400\_err} & \multirow{2}{*}{\parbox{3.8in}{Time of arrival error with reference to 400.1953125 MHz for the specific sub-burst. }} \\
    & & &\\
    50 & MJD & \texttt{mjd\_inf} & \multirow{2}{*}{\parbox{3.8in}{Time of arrival with reference to infinite frequency for the specific sub-burst. }} \\
    & & &\\
    51 & MJD & \texttt{mjd\_inf\_err} & \multirow{2}{*}{\parbox{3.8in}{Time of arrival error with reference to infinite frequency for the specific sub-burst. }} \\
    & & &\\
    52 & s & \texttt{width\_fitb} & Width of sub-burst using \fitburst \\
    53 & s & \texttt{width\_fitb\_err} & Width error of sub-burst using \fitburst \\
    54 &... & \texttt{sp\_idx} & Spectral index for the sub-burst \\
    55 & ...& \texttt{sp\_idx\_err} & Spectral index error for the sub-burst \\
    56 &... & \texttt{sp\_run} & Spectral running for the sub-burst \\
    57 &...& \texttt{sp\_run\_err} & Spectral running error for the sub-burst \\
    58 & MHz & \texttt{high\_freq} &\multirow{2}{*}{\parbox{3.8in}{Highest frequency band of detection for the sub-burst at FWTM}} \\
    & & &\\
    59 & MHz & \texttt{low\_freq} & \multirow{2}{*}{\parbox{3.8in}{Lowest frequency band of detection for the sub-burst at FWTM}} \\
    & & &\\
    60 & MHz & \texttt{peak\_freq} & \multirow{2}{*}{\parbox{3.8in}{Peak frequency of detection for the sub-burst}} \\
    & & &\\
    61 & ... & \texttt{chi\_sq} & Chi-squared from \fitburst \\
    62 & ... & \texttt{dof} & Number of degrees of freedom in \fitburst  \\
    63 & ... & \texttt{flag\_frac} & Fraction of spectral channels flagged in \fitburst \\
    64 & ... & \texttt{r\_cc} & Contamination rate (see Section~\ref{sec:pcc}). \\
     & & &\\
     & & &\\
    \hline
    \\[-2ex]
    \label{ta:catalog descriptions}
    \end{longtable}
    \end{ThreePartTable}

\section{Additional Figures}
\label{sec:extrafigs}

\begin{figure}[ht]
    \centering
    \includegraphics[width=\textwidth]{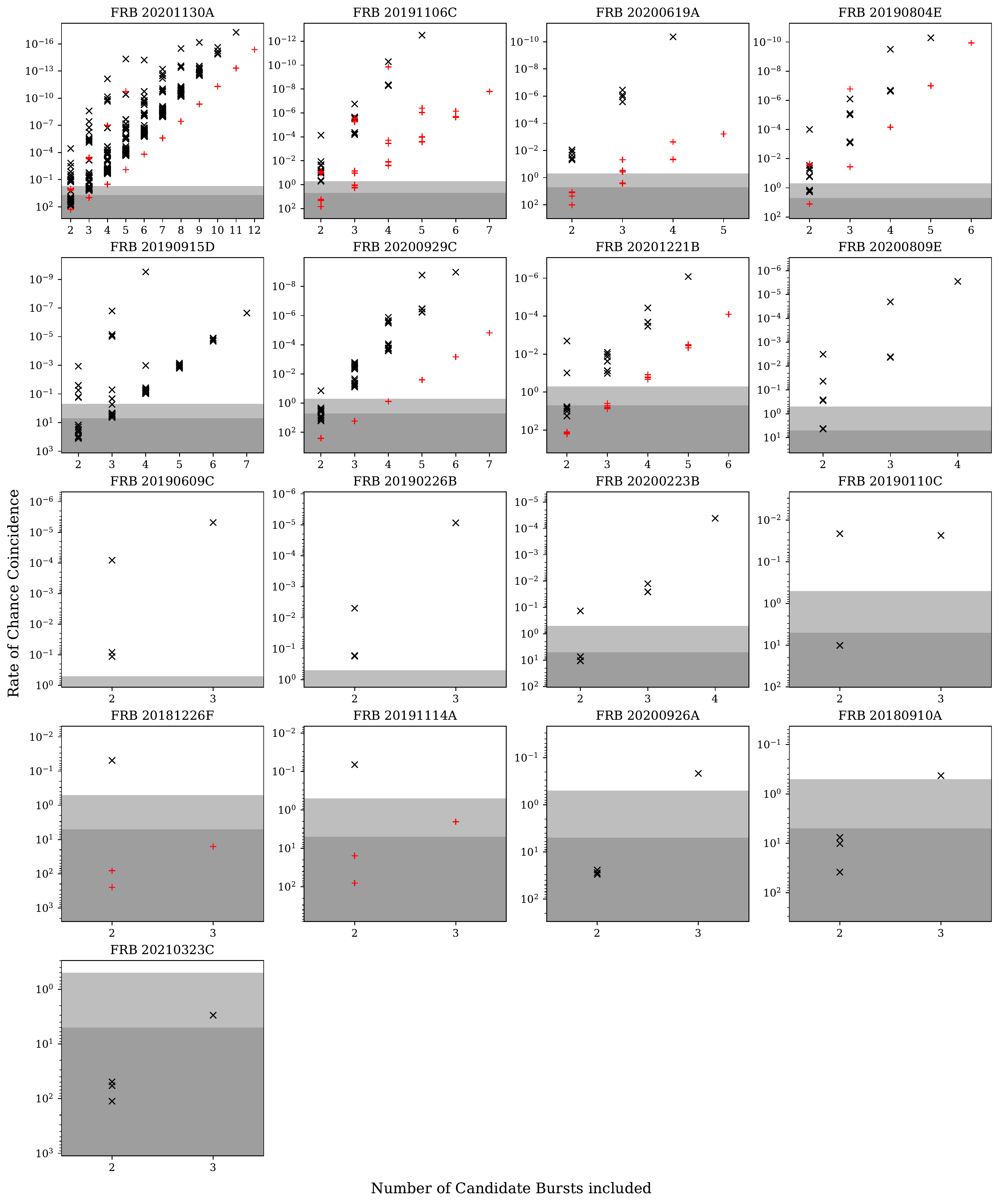}
    \caption{Contamination rate ($R_\mathrm{cc}$) of all combinations of bursts in clusters of 3+ events. Note that the vertical axes are inverted. The gray regions indicate $0.5 \leq R_\mathrm{cc} < 5$ (silver sample; lighter region) and $R_\mathrm{cc} \geq 5$ (darker region). Potential outlier events are marked by red pluses, all other events by black crosses (\S\ref{sec:pcc}).}
    \label{fig:outliers}
\end{figure}

\begin{figure}[ht]
    \centering
    \includegraphics[width=0.9\textwidth]{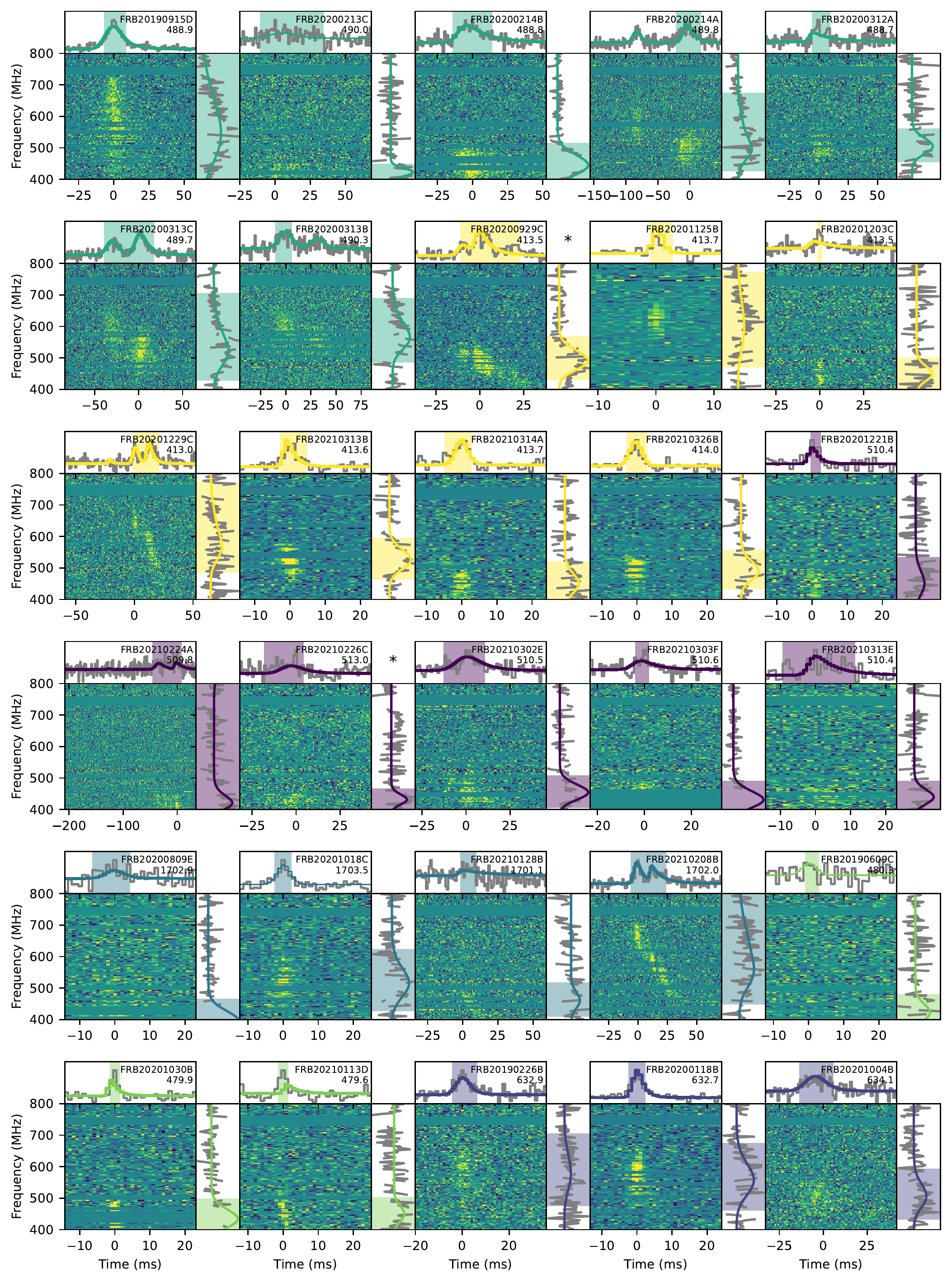}
    \caption{Continued from Figure~\ref{fig:gold_wfalls_0}.}
    \label{fig:gold_wfalls_1}
\end{figure}

\begin{figure}[ht]
    \centering
    \includegraphics[width=0.9\textwidth]{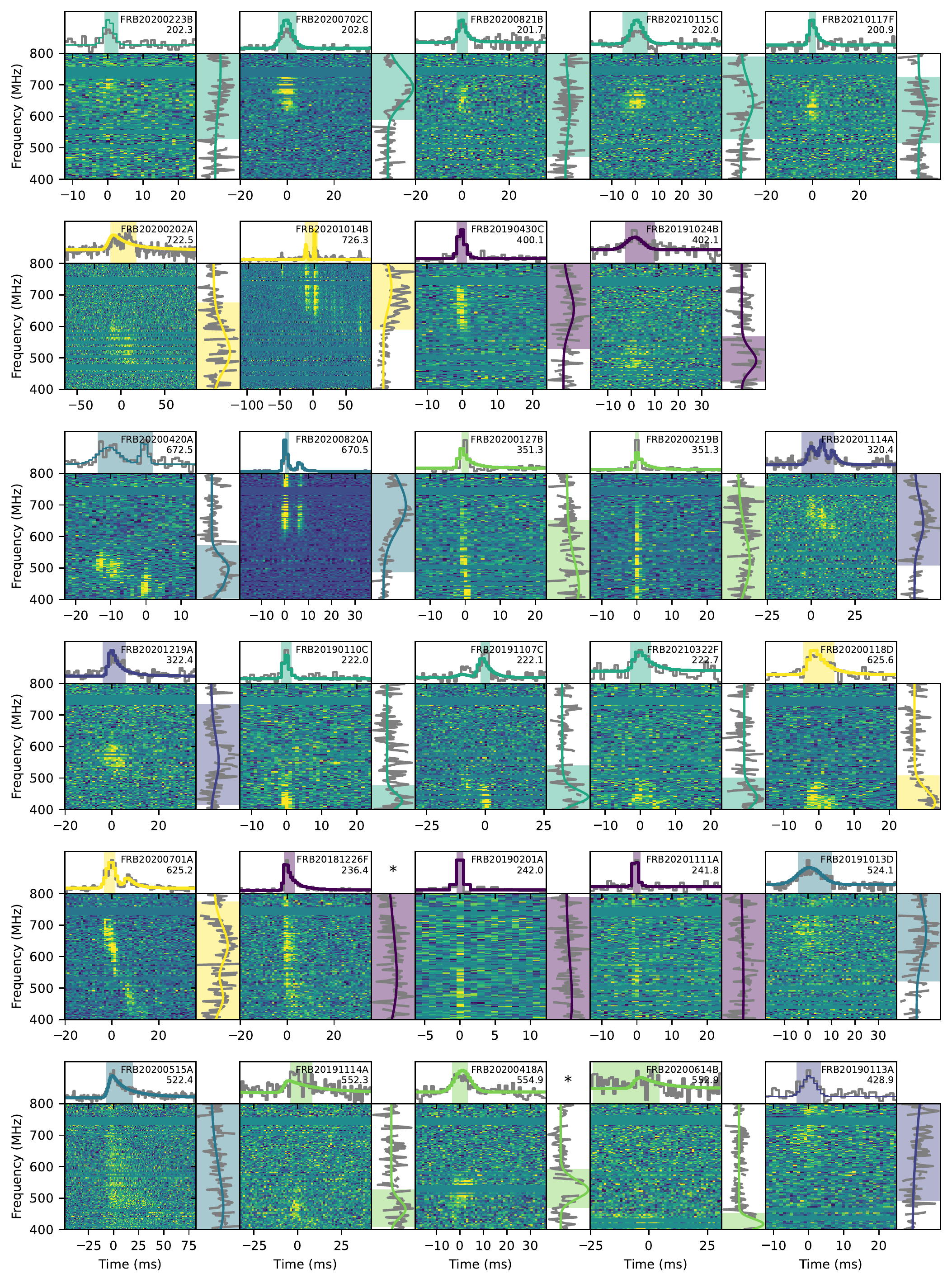}
    \caption{Continued from Figure~\ref{fig:gold_wfalls_0}.}
    \label{fig:gold_wfalls_2}
\end{figure}

\begin{figure}[ht]
    \centering
    \includegraphics[width=0.9\textwidth]{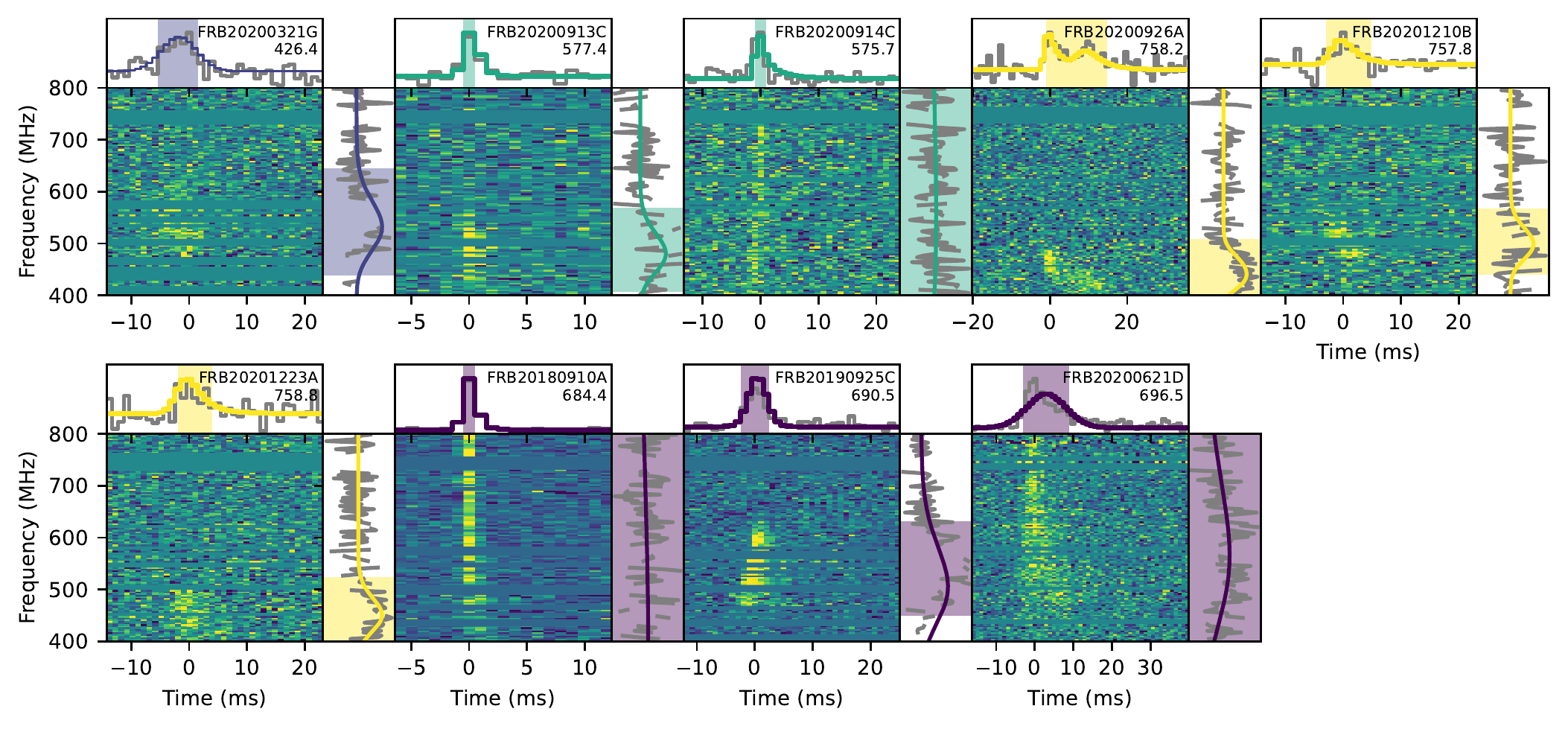}
    \caption{Continued from Figure~\ref{fig:gold_wfalls_0}.}
    \label{fig:gold_wfalls_3}
\end{figure}

\end{document}